\documentclass[12pt]{article}

\usepackage{verbatim,color,amssymb}
\usepackage{amsmath}					
\usepackage{amsthm}					
\usepackage{tikz}
\usepackage{natbib}
\usepackage{multirow}
\usepackage{setspace}
\usepackage[mathscr]{euscript}
\usepackage{fancyhdr}
\usepackage{enumitem}
\usepackage{graphicx}
\usepackage{geometry}
\usepackage[normalem]{ulem}
\usepackage[bookmarks=false]{hyperref}

\usepackage{tabularx, booktabs}
\newcolumntype{Y}{>{\centering\arraybackslash}X}

\usepackage{lscape}

\usepackage{subfig}
\usepackage{caption}
\usepackage{float}
\usepackage{lineno}

\def\ttd{\texttt{d}}
\def\ttm{\texttt{m}}
\def\tts{\texttt{s}}
\def\ttu{\texttt{u}}

\def\C{{\cal C}}

\def\X{{\cal X}}

\def\Y{{\cal Y}}

\def\wh{\widehat}
\def\wt{\widetilde}

\def\log{\hbox{log}}

\def\Bern{\hbox{Bernoulli}}
\def\Beta{\hbox{Beta}}

\def\Dir{\hbox{Dir}}

\def\Ga{\hbox{Ga}}

\def\Mult{\hbox{Mult}}

\def\P_25_ICML{{\it Proceedings of the 25th international conference on Machine learning}}

\def\refhg{\hangindent=20pt\hangafter=1}
\def\refmark{\par\vskip 2mm\noindent\refhg}

\def\refhg{\hangindent=20pt\hangafter=1}
\def\refmark{\par\vskip 2mm\noindent\refhg}

\def\bse{\begin{eqnarray*}}
\def\ese{\end{eqnarray*}}
\def\be{\begin{eqnarray}}
\def\ee{\end{eqnarray}}
\def\bq{\begin{equation}}
\def\eq{\end{equation}}

\def\wh{\widehat}

\def\trans{^{\rm T}}
\def\data{{\hbox{data}}}

\def\th{^{th}}

\def\b1e{{\mathbf e}}
\def\b1f{{\mathbf f}}

\def\bk{{\mathbf k}}

\def\bP{{\mathbf P}}

\def\by{{\mathbf y}}

\def\bz{{\mathbf z}}

\newcommand{\bmu}{\mbox{\boldmath $\mu$}}

\newcommand{\bpi}{\mbox{\boldmath $\pi$}}

\newcommand{\btheta}{\mbox{\boldmath $\theta$}}

\newcommand{\bzeta}{\mbox{\boldmath $\zeta$}}

\newcommand{\blambda}{\mbox{\boldmath $\lambda$}}

\newcommand{\btau}{\mbox{\boldmath $\tau$}}

\renewcommand\footnoterule{\kern-3pt \hrule \textwidth 2in \kern 2.6pt}
\newcommand*{\myalign}[2]{\multicolumn{1}{#1}{#2}}

\def\boxit#1{\vbox{\hrule\hbox{\vrule\kern6pt \vbox{\kern6pt \textcolor{blue}{#1}\kern6pt}\kern6pt\vrule}\hrule}}

\def\authorfootnote#1{{\let\thefootnote\relax\footnotetext{#1}}}

\graphicspath{{figures/}}
\usepackage[ruled,vlined]{algorithm2e}

\allowdisplaybreaks

\begin{document}
\thispagestyle{empty}
\baselineskip=28pt

\begin{center}
{\LARGE{\bf Bayesian Semiparametric \\ Markov 
Renewal Mixed Models 
for Vocalization Syntax
}}
\end{center}
\baselineskip=12pt

\vskip 20pt 
\begin{center}
Yutong Wu\\
Department of Mechanical Engineering\\
The University of Texas at Austin\\
204 E. Dean Keeton Street C2200, Austin, TX 78712-1591, USA\\
yutong.wu@utexas.edu\\
\vskip 10pt 
Erich D. Jarvis\\
The Rockefeller University, New York, NY 10065, USA\\
Howard Hughes Medical Institute, Chevy Chase, MD 20815, USA\\
ejarvis@rockefeller.edu \\
\vskip 10pt 
Abhra Sarkar\\
Department of Statistics and Data Sciences\\ 
The University of Texas at Austin\\ 
2317 Speedway D9800, Austin, TX 78712-1823, USA\\
abhra.sarkar@utexas.edu\\ 
\vskip 10pt 
\end{center}

\vskip 20pt 
\begin{center}
{\Large{\bf Abstract}} 
\end{center}
\baselineskip=12pt

{Speech and language play an important role in human vocal communication. 
Studies have shown that vocal disorders can result from genetic factors.  
In the absence of high-quality data on humans,  mouse vocalization experiments in laboratory settings have been proven useful in providing valuable insights into mammalian vocal development,  including especially the impact of certain genetic mutations.
Data sets from mouse vocalization experiments usually consist of categorical syllable sequences along with continuous inter-syllable interval times (ISIs) for mice of different genotypes vocalizing under various contexts.  
ISIs are of particular importance as increased ISIs can be an indication of possible vocal impairment. 
Statistical methods for properly analyzing ISIs along with the transition probabilities have however been lacking. 
In this paper, we propose a class of novel Markov renewal mixed models 
that capture the stochastic dynamics of both state transitions and ISI lengths.  
Specifically,  we model the transition dynamics and the ISIs using Dirichlet and gamma mixtures, respectively,  allowing the mixture probabilities in both cases to vary flexibly with fixed covariate effects as well as random individual-specific effects.
We apply our model to analyze the impact of a mutation in the Foxp2 gene on mouse vocal behavior. 
We find that genotypes and social contexts significantly affect the length of ISIs but, compared to previous analyses,  the influences of genotype and social context on the syllable transition dynamics are weaker.}

\vskip 8mm
\baselineskip=12pt
\noindent\underline{\bf Some Key Words}: 
Clustering,
Dirichlet mixtures, 
Gamma mixtures, 
Markov renewal processes,
Mixed effects models, 
Mouse vocalization experiments
 
\par\medskip\noindent
\underline{\bf Short Title}: Markov Renewal Mixed Models for Vocalization Syntax

\par\medskip\noindent
\underline{\bf Corresponding Author}: Abhra Sarkar (abhra.sarkar@utexas.edu)

\pagenumbering{arabic}
\setcounter{page}{0}
\newlength{\gnat}
\setlength{\gnat}{16pt}
\baselineskip=\gnat

\newpage

\section{Introduction}\label{sec: intro}

Spoken language plays a crucial role in almost every aspect of human life as we use it to share information, communicate ideas, and express emotions. 
However, our vocal behaviors might be restrained by a wide variety of impairments, some of which are highly inheritable. 
{According to the National Institute on Deafness and Other Communication Disorders,  the prevalence of speech and sound disorder among young children is 8 to 9 percent and the majority of such disorders have no known cause \citep{NIH-NIDCD:2020}.   
As many speech-related disorders are inheritable \citep{vargha1998neural}, } 
it is thus important to study the genetic and evolutionary development of human vocal communication to identify and remedy vocal disorders.

Since data on human vocalization disorders are not easily available,  
neither can humans be studied under experimentally induced impairing conditions, 
neuroscientists have turned to studying mouse vocalization systems to gain insights into human vocal communication processes  \citep{jarvis2019evolution}.
Although unlike speech, the mouse vocalization is mostly innate \citep{arriaga2013mouse, mooney2020neurobiology} and is a particularly attractive model to study for several reasons. 
Adult mice `sing' ultrasonic vocalizations (USVs) to communicate with each other. 
Being mammals, they are also physiologically and genetically similar to us humans.
The patterns of USVs may be influenced by the mouse genotype or environmental factors such as stimulating social contexts \citep{Chabout_etal:2015}. 
Mouse vocalization data sets thus typically comprise songs sung by mice from different genotypes under various social contexts. 
Systematic differences in the syllable dynamics across various genotypes and social conditions 
can provide insights into their roles on vocal abilities and behavior \citep{Chabout_etal:2016}.

Our research is motivated by the need for sophisticated statistical methods for analyzing mouse vocalization syntax 
generated in laboratory experiments that are conducted to understand the effects of certain genetic mutations and social contexts on mouse vocal behavior.
We introduce a novel class of Bayesian mixed models for
analyzing categorical sequences 
with continuous inter-state interval times 
under the influence of multiple exogenous factors.   
In particular, the values of the exogenous factors remain fixed throughout the sequence and contributes a fixed group effect. 
Each sequence is also associated with an individual that exhibits a random individual effect. 
We are interested in the inference of the stochastic dynamics of the sequences, 
specifically, the transition dynamics of the discrete states as well as the distribution of the continuous inter-state interval times. 
Statistical methods for analyzing the syllable dynamics have previously been developed by \cite{Holy_Guo:2005} and \cite{Chabout_etal:2015,Chabout_etal:2016}. 
\cite{sarkar2018bayesian} developed a flexible Bayesian mixed effects Markov model for vocalization syntax 
incorporating exogenous influences of covariates as well as random heterogeneity of the sampled mice. 
While these methods looked in detail into the systematic differences in the syllable dynamics, 
they did not properly analyze the inter-syllable intervals (ISIs)
which can be an additional important indicator of vocal deficits 
as impaired mice will tend to remain  silent with longer ISIs. 
To accommodate this effect, 
the aforementioned methods {discretized} large ISIs {into one or multiple} special `silent' syllables 
and then treated the songs as Markov sequences with this appended vocabulary. 
This practice, however, mixes the influences of covariates on transition probabilities and ISIs which could result in less accurate scientific conclusions. 
The previous methods
largely ignore the differences in the distributions of the ISIs from different mice under various combinations of the covariate values 
and may miss out important evidence that can be deduced by properly modeling the ISIs.

Here we develop an approach to address these concerns by appropriately analyzing 
the syllable transition dynamics using {a slightly modified version of the} mixed Markov model of \cite{sarkar2018bayesian}
while separately modeling the distribution of ISIs using a novel flexible mixed model of gamma mixtures, 
thereby providing inference for both syllable transitions dynamics and their ISIs. 
The method accommodates fixed covariate effects as well as random individual effects in both the syllable transition dynamics and the ISI distributions. 
A hierarchical cluster inducing mechanism for the levels of the covariates allows straightforward, formal tests of their significance.  
We design an efficient Markov chain Monte Carlo sampler for fitting our model. 
We demonstrate the performance of our model by analyzing a data set 
where the mice are either wild-type or carry a mutation on the Foxp2 gene implicated in causing vocal impairment in humans.
Previous analyses of this data set by \cite{Chabout_etal:2015,Chabout_etal:2016,sarkar2018bayesian}, 
with large ISIs treated as an artificial syllable, have shown significant differences in the syllable dynamics between different genotypes and social contexts. 
When reanalyzed using our proposed approach, 
the results suggest that genotype and context strongly impact the ISIs, 
but their influences on the syllable transition dynamics are weaker than what previous analyses	 had inferred.

Our proposed model is a type of Markov renewal process (MRP) where 
the state transitions evolve as a Markov chain 
while the state durations follow a transition density function that depends on {both the previous and the current} state. 
MRPs were originally introduced  by \cite{pyke1961markov}. 
MRPs and their variations, including semi-Markov models \citep{levy1954processus,smith1955regenerative}, have found success in a variety of applications 
such as in modeling clinical trials \citep{weiss1965semi}, 
sleeping patterns \citep{yang1973use}, HIV disease occurrences \citep{foucher2005semi}, etc.
Bayesian methods for MRPs have also been developed.
\cite{phelan1990bayes} designed an MRP
where the prior consisted of
a family of Dirichlet distributions for transition matrices 
and a Beta family of Levy processes for state duration times. 
\cite{muliere2003reinforced} and \cite{bulla2007bayesian} developed Bayesian nonparametric reinforced MRPs. 
Bayesian MRPs with Weibull distributed inter-occurrence times have been developed for seismic data in \cite{alvarez2005estimation, epifani2014bayesian}. 
{Unlike most classical MRPs that focus on modeling a single sequence,  we jointly model a collection of sequences, 
each one associated with an individual as well as a set of time-invariant external covariates, 
accommodating fixed effects of the covariates as well as random effects of the subjects 
for both the transition probability matrices and the distribution of the ISIs. } 
We also allow the selection of important covariates for both the state transition dynamics and the ISI distribution via probabilistic partitioning of the covariate levels. 
These are in contrast to existing methods on classical MRPs 
where typically only a single sequence is modeled and the distribution of state duration depends only on the current state.

{Besides being directly useful for analyzing mouse vocalization data sets, the methodology proposed in the paper can be used for a great variety of applications where MRPs are useful {such as the examples cited above}. 
Our proposed model for the ISIs may also be of independent statistical interest in developing mixture models for continuous variables with mixed covariate and individual effects 
which, to our knowledge, have not been explored much in the literature.}

The remainder of this paper is organized as follows. 
In Section \ref{sec: data set and background}, we provide details of the Foxp2 data set and its scientific background, and review some previous statistical methods. 
In Section \ref{sec: mrmm}, we present our novel Bayesian Markov renewal mixed model.  
In Section \ref{sec: hyper-parameters post inf}, we briefly outline our Markov chain Monte Carlo (MCMC) algorithm to sample from the posterior. 
We illustrate the results of our method applied to the Foxp2 data set in Section  \ref{sec: real data results}.   
Section \ref{sec: discussion} contains discussions and  concluding remarks. 
\section{Data Set and Preliminaries}\label{sec: data set and background}

The Foxp2 (foxhead-box P2) gene is a transcription factor that regulates other genes \citep{Chabout_etal:2016}. 
It is found in mice and in similar forms in humans \citep{fisher2009foxp2},  
and its mutation has been implicated to cause speech and language deficits in adults \citep{lai2001forkhead}. 
Previously, \cite{Fujita_etal:2008} compared the vocalizations for wild-type, heterozygous Foxp2 and homozygous Foxp2 mice 
and showed that both heterozygous and homozygous Foxp2 mice have vocal impairment to some extent. 
\cite{Castellucci_etal:2016} showed that mice with a Foxp2 mutation vocalize less and produce shorter syllable sequences. 
\cite{Gaub_etal:2016} discovered that compared to wild-type, mice with Foxp2 mutations displayed quantitative differences in USVs.
Other than genotypes, social contexts can also influence mouse vocalization. 
\cite{Chabout_etal:2012} showed that 
the amount of USVs emitted by male mice is positively correlated with the scale of their social interactions. 
\cite{Gaub_etal:2016} studied the USVs of adult mice with increasing stimulus intensity including water and female urine.

\begin{figure}[ht]
\centering
\subfloat[Syllable types.]{
\includegraphics[height=4cm, width=7cm, trim=0cm 0.25cm 0cm 0cm, clip=true]{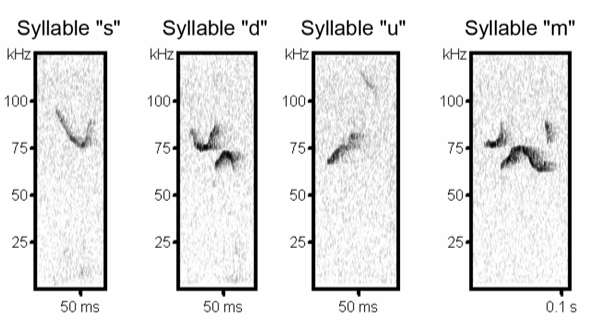}
\label{fig: syllables}}\\
\subfloat[Part of a song produced by a wild-type male mouse under the urine context ($U$).]{
\includegraphics[height=3cm, width=15.5cm, trim=0cm 0.5cm 0cm 0.5cm, clip=true]{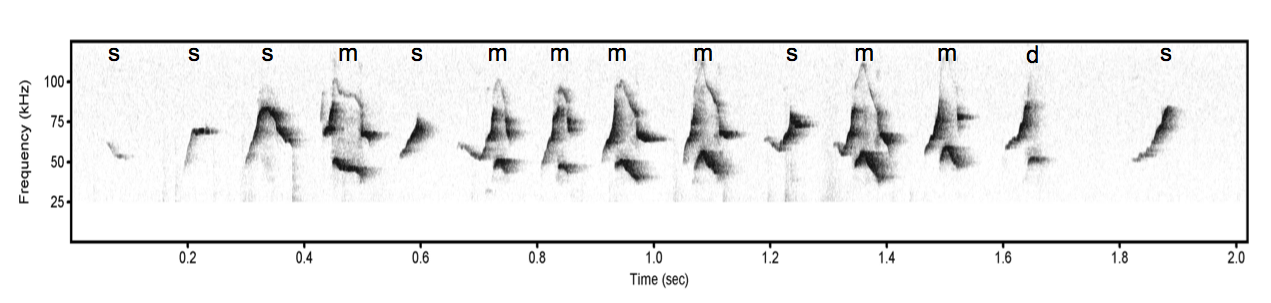}
\label{fig: example song}}
\caption{Spectral diagrams of mouse vocalizations, reproduced from \citealp{Chabout_etal:2016} with permission.}
\label{fig:spectral}
\end{figure}

\begin{figure}[ht]
\centering
\includegraphics[width=0.6\textwidth]{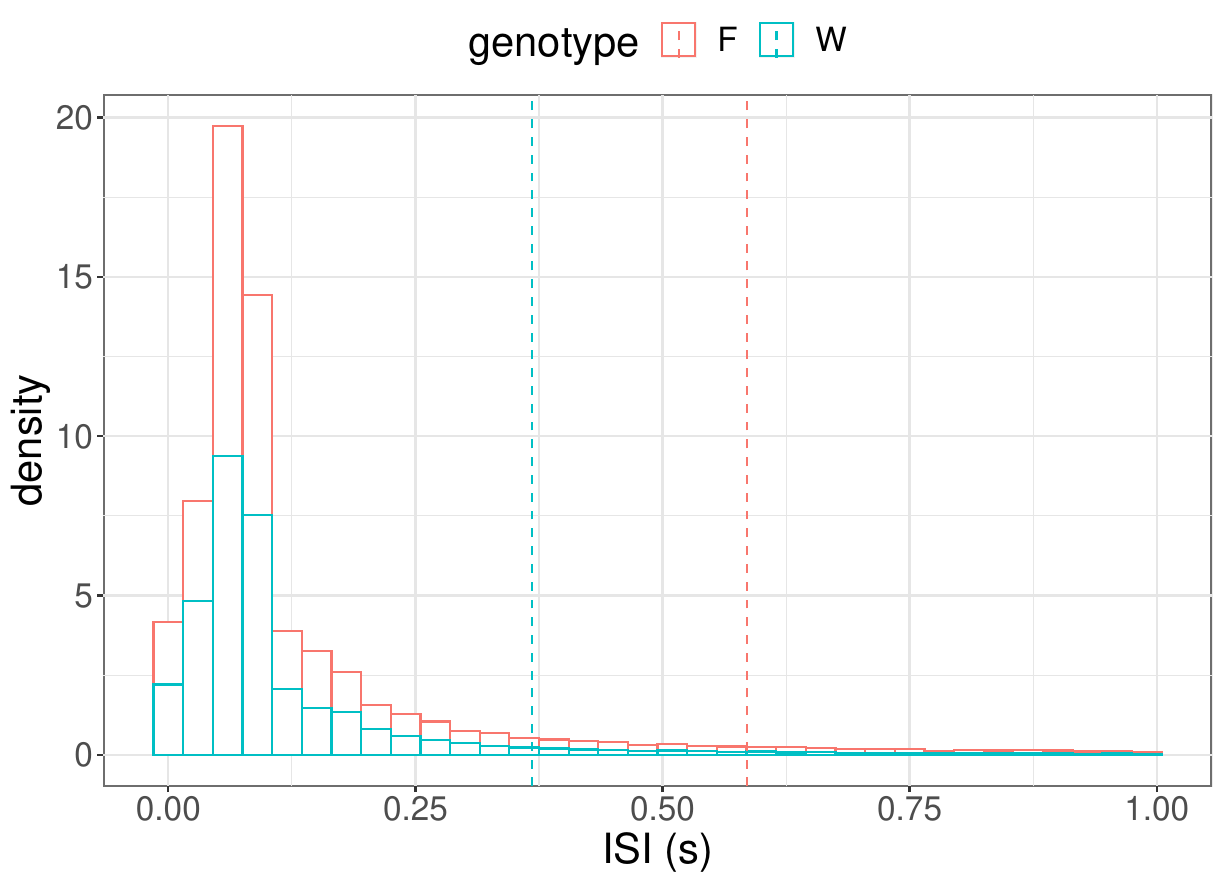}
\caption{Histogram of the ISIs grouped by genotypes  {with the group means (dotted lines)}.   
The histogram has a long tail 
and the x-axis is cut off at 1 second for better visualization.  
The actual range of the ISIs is from 0.01 seconds to 258.8 seconds.}
\label{fig:isi_hist}
\end{figure}

The overall properties of the song syllables provides some information.
The sequencing rules of the syllables, call syntax, provides additional information about song complexity.
Research has thus also been conducted on the vocabulary and structure of the songs, referred to as `syntax' 
\citep{Holy_Guo:2005,Moles_etal:2007,Scattoni_etal:2011,Musolf_etal:2015,Gaub_etal:2016}. 
Adopting the vocabulary in \cite{Chabout_etal:2012,Chabout_etal:2015,Chabout_etal:2016}, 
the mouse vocalization syllables can be grouped into four categories based on their spectral features (Figure~\ref{fig: syllables}) as:
 `\texttt{s}': simple syllables without any pitch jumps;  `\texttt{u}': complex syllables with a single upward pitch jump; `\texttt{d}': complex syllables with a single downward pitch jump; and  `\texttt{m}': more complex syllables with a series of multiple pitch jumps.
Songs are made of a sequential arrangements of these syllables (Figure~\ref{fig: example song}) 
with 
ISIs varying mostly between 0 and 250 milliseconds (Table~\ref{tab: data under context A}, Figure~\ref{fig:isi_hist}).

\begin{table}[ht]
\centering
\begin{tabularx}{0.8\textwidth}{c *{4}{Y}}
\cline{1-4}
Mouse ID & Genotype & Syllable & ISI (in seconds)\\ \hline 
1 & $F$ & \texttt{s} & $0.082$ \\ 
1 & $F$ & \texttt{s} & $0.017$ \\ 
1 & $F$ & \texttt{m} & $0.114$ \\ 
$\vdots$ & $\vdots$ & $\vdots$ & $\vdots$ \\
2 & $W$ & \texttt{s} & $1.546$  \\ 
2 & $W$ & \texttt{d} & $0.712$ \\ 
2 & $W$ & \texttt{s} & $0.549$ \\
\hline
\end{tabularx}
\caption{Part of the data set associated with the anesthetized female context ($A$).}
\label{tab: data under context A}
\end{table}

The Foxp2 data set that we analyze here collects the songs produced by mice of wild-type ($W$) and mice that have a mutation on the Foxp2 gene ($F$) under three different social contexts -- fresh female urine on a cotton tip placed inside the male's cage ($U$), 
awake and behaving adult female placed inside the cage ($L$), 
and one anesthetized female placed on the lid of the cage ($A$). 
{The data set has $70818$ rows, including $49$ songs sung by $18$ mice, $10$ with the Foxp2 mutation and $8$ wild-types. 
Five mice sang 2 songs each (with id 1, 2, 4, 10, and 11) and the rest $13$ mice sang 3 songs each.  
Each mouse sang the songs under different contexts. 
Therefore, the combination of covariates and individuals associated with each song is actually unique.  
The distribution of songs across different combinations of covariates is presented in Table~\ref{tab: Foxp2 empirical} (left). 
There is no missing or censored data.
The empirical distribution of syllable transitions is displayed in Table~\ref{tab: Foxp2 empirical} (right). 
The number of transition types stratified by genotypes and social contexts can be found in the Appendix A in the supplementary materials. }

\begin{table}[ht]
\parbox{.4\linewidth}{
\centering
\caption*{Distribution of songs}
\vspace{-2ex}
\begin{tabular}{*{2}cc}
Genotype & Context & \# Songs \\ \hline
$F$                   & $U$         & 8 		 \\
 $F$                     & $L$          & 10  	\\
 $F$                 & $A$         & 10  \\
$W$                   & $U$         & 6 		 \\
 $W$                     & $L$          & 8 	\\
 $W$                 & $A$         & 7  \\\hline
\end{tabular}
}
\quad~~~
\parbox{.5\linewidth}{
\centering
\caption*{Distribution of syllable transitions}
\vspace{-2ex}
\begin{tabular}{cc*{4}r}
\multicolumn{1}{c}{}  & \multicolumn{1}{c}{} & \multicolumn{4}{c}{current syllable}\\
\multicolumn{1}{c}{}  & \multicolumn{1}{c}{} & \myalign{c}{\texttt{d}} & \myalign{c}{\texttt{m}} & \myalign{c}{\texttt{s}} & \myalign{c}{\texttt{u}}  \\ \cline{2-6}
&\texttt{d} 	& 2780          & 964       & 5980           & 268        	 \\
preceding &\texttt{m} 	& 987          & 983        & 3187           & 257          	\\
syllable &\texttt{s} 	& 5920         & 3213       & 42138          & 1742       \\
&\texttt{u} 	& 305          & 257        & 1705            & 132          		\\\cline{2-6}
\end{tabular}
}
\caption{Empirical distributions of the Foxp2 data set: song distribution across different combinations of covariates (left) and number of transitions for all pairs of syllables (right). }
\label{tab: Foxp2 empirical}
\end{table}

Since the complexity of the four syllables vary, 
it is {reasonable to assume} that mice with vocal impairments will produce songs with fewer transitions to difficult syllables such as \texttt{m}. 
It is also expected that mice with vocal impairments  tend to remain more silent with longer ISIs. 
{Table~\ref{tab: isi mean} shows the empirical distribution of the mean and standard deviation of ISIs grouped by each covariate and
Figure~\ref{fig:isi_hist} provides the histograms of ISIs given a genotype.  
We see that mice with the Foxp2 mutation are likely to have longer ISI than that of wild-types,  with a 42\% increase in the empirical distribution of ISI mean.
}

\begin{table}[ht]
\parbox{.44\linewidth}{
\centering
\begin{tabular}{ccc}
 \multicolumn{1}{c}{} & mean & s.d.  \\ \hline
$F$ 	& 0.228 (0.227) & 0.433 (0.417) \\
$W$ 	& 0.161 (0.163) & 0.333 (0.320)   	\\
\hline
\end{tabular}
}
\parbox{.55\linewidth}{
\centering
\begin{tabular}{ccc}
 \multicolumn{1}{c}{} & mean & s.d.  \\ \hline
$U$ 	& 0.257 (0.237) & 0.504 (0.424) \\
$L$ 	& 0.166 (0.168) & 0.319 (0.335)   	\\
$A$ 	& 0.211 (0.218) & 0.426 (0.399)   	\\
\hline
\end{tabular}
}\\[8pt]
\hspace{37pt}
\parbox{.44\linewidth}{
\centering
\begin{tabular}{ccc}
 \multicolumn{1}{c}{} & mean & s.d.  \\ \hline
(\ttd,\ttd) & 0.134 (0.127) & 0.236 (0.241)  \\ 
(\ttd,\ttm) & 0.119 (0.126) & 0.261 (0.239) \\ 
(\ttd,\tts) & 0.180 (0.181) &  0.355 (0.351)  \\ 
(\ttd,\ttu) & 0.172 (0.144) & 0.343 (0.300) \\ 
(\ttm,\ttd) & 0.118 (0.138) &  0.234 (0.267)  \\ 
(\ttm,\ttm) & 0.120 (0.129) &  0.231 (0.245)  \\ 
(\ttm,\tts) & 0.145 (0.144) &  0.295 (0.301) \\ 
(\ttm,\ttu) & 0.120 (0.155) &  0.228 (0.305) \\ 
\hline
\end{tabular}
}
\parbox{.55\linewidth}{
\centering
\begin{tabular}{ccc}
 \multicolumn{1}{c}{} & mean & s.d.  \\ \hline
(\tts,\ttd) & 0.168 (0.154) &  0.341 (0.303)  \\ 
(\tts,\ttm) & 0.131 (0.129) &  0.253 (0.246)  \\ 
(\tts,\tts) & 0.217 (0.221) & 0.423 (0.407) \\ 
(\tts,\ttu) & 0.167 (0.151) &  0.338 (0.294)  \\ 
(\ttu,\ttd) & 0.130 (0.157) &  0.290 (0.319) \\ 
(\ttu,\ttm) & 0.126 (0.130) &  0.210 (0.246) \\ 
(\ttu,\tts) & 0.165 (0.144) &  0.373 (0.303)  \\ 
(\ttu,\ttu) & 0.139 (0.174) &  0.294 (0.341) \\
\hline
\end{tabular}
}
\caption{{The empirical and posterior (in parentheses) means and standard deviations of the ISIs, grouped by different covariate levels.}}
\label{tab: isi mean}
\end{table}

The development of sophisticated statistical methods for mouse vocalization syntax started with \cite{Holy_Guo:2005} 
who analyzed the songs using a Markov model. 
\cite{Chabout_etal:2015,Chabout_etal:2016} developed statistical tests for accessing global and local syntax differences across genotypes and social contexts. 
\cite{sarkar2018bayesian} developed a mixed Markov model for the transitions of four syllables and an extra artificial syllable for large ISIs.
In the latter three works, each ISI of length greater than 250 milliseconds (ms) was treated as a silent state (`\texttt{x}'). 
Inference was then performed  
treating the resulting sequences as Markov chains with five states \{\texttt{d}, \texttt{m}, \texttt{s}, \texttt{u}, \texttt{x}\}.
Though this was done to achieve significant analytical convenience, the resulting stochastic dynamics also got dominated by transitions to \texttt{x}.
Moreover, the distribution of the ISIs varied greatly between different experimental conditions and subjects (Figure \ref{fig:isi_hist}). 
Ignoring the ISIs with lengths shorter than 250 ms, 
as well as treating longer ISIs as blocks of silent syllables, 
resulted in loss of important information 
in addition to diluting the transition dynamics among the original syllables.
In order to obtain a more accurate inference of the data set, 
it is important to treat the ISIs differently from the four original syllables and model them properly as a continuous variable.

Our proposed approach addresses these concerns.
We model the ISIs separately instead of treating it ad-hocly as the `silence' syllable.
In this way, the ISIs can be used as evidence for vocal impairment aside from the transitions of the four syllables.
Moreover, we allow both the transition probabilities and the mixture probabilities for the ISIs to be governed by 
a convex combination of population level fixed effects and individual level random effects.  
\section{Markov Renewal Mixed Models}\label{sec: mrmm}

In this section, we detail our model for mouse vocalization syntax. 
Consider a sequence $s$ of $T_s$ syllables. 
$y_{s,t}$ denotes the syllable at time $t$ for sequence $s$, and is one of $\Y=\{\texttt{d},\texttt{m},\texttt{s},\texttt{u}\}=\{1,2,3,4\}$. 
The collection of syllables is denoted by $\{y_{s,t}\}_{s=1,t=1}^{s_{0},T_{s}}$ where $s_0$ is the total number of sequences.
Within a sequence $s$, we have $T_s-1$ inter-syllable interval (ISI) times, 
denoted by {$\{\tau_{s,t}\}_{s=1,t=2}^{s_0,T_s}$,} 
where $\tau_{s,t}$ represents the interval time between the {$(t-1)\th$ and $t\th$} syllable for sequence $s$. 
Each sequence $s$ is generated under two exogenous factors --  
genotype $x_{s,1}\in \X_1=\{F,W\}=\{1,2\}$,  
and social context $x_{s,2}\in \X_2=\{U,L,A\}=\{1,2,3\}$, as described in Section~\ref{sec: data set and background}. 
With some abuse, 
we use the same notation to denote the variables as well as their specific values, greatly simplifying the exposition. 
Table \ref{tab: notation} provides a complete list of variables used in our model detailed below.

\begin{figure}[ht!]
    \centering
\begin{tikzpicture}
\draw (-4.5,0) circle (13pt);
\node at (-4.5,0) {$i_{s}$};

\draw (-3,0) circle (13pt);
\node at (-3,0) {$x_{s,1}$};
\node at (-2.25,0) {$\dots$};
\draw (-1.5,0) circle (13pt);
\node at (-1.5,0) {$x_{s,p}$};

\draw[rounded corners] (-0.7, -0.7) rectangle (10.7, 0.7) {};
\draw[rounded corners] (-3.7, -0.7) rectangle (-0.8, 0.7) {};

\draw (0,0) circle (13pt);
\draw (2,0) circle (13pt);
\draw (6,0) circle (13pt);
\node at (0,0) {$y_{s,1}$};
\node at (2,0) {$y_{s,2}$};
\node at (6,0) {$y_{s,t}$};
\draw[->,thick] (0.5,0) -- (1.5,0); 
\draw[->,thick] (2.5,0) -- (3.5,0);
\draw[dotted,thick] (3.6,0) -- (4.4,0);
\draw[->,thick] (4.5,0) -- (5.5,0);
\draw[->,thick] (6.5,0) -- (7.5,0);
\draw[dotted,thick] (7.6,0) -- (8.4,0);
\node at (1,0.3) {$\tau_{s,2}$};
\node at (3,0.3) {$\tau_{s,3}$};
\node at (5,0.3) {$\tau_{s,t}$};
\node at (7,0.3) {$\tau_{s,t+1}$};
\node at (6,0) {$y_{s,t}$};
\draw[->,thick] (8.5,0) -- (9.5,0);
\node at (9,0.3) {$\tau_{s,T_{s}}$};
\node at (10,0) {$y_{s,T_{s}}$};
\draw (10,0) circle (13pt);
\end{tikzpicture}
    \caption{{Graphical model showing the data structures: $y_{s,t}$ denotes the observed state at the $t\th$ time location in the $s\th$ sequence and  
    $\tau_{s,t}$ denotes the observed ISIs between the states $y_{s,t-1}$ and $y_{s,t}$; 
    each sequence $s$ is also associated with an individual $i_{s}$ and a set of exogenous time-invariant covariates $x_{s,1},\dots, x_{s,p}$. 
    The Markov renewal mixed model considered in this article analyzes the state transitions $y_{s,t}$ and the {ISI lengths} $\tau_{s,t}$,  accommodating fixed effects of the covariates $x_{s,1},\dots, x_{s,p}$ and random effects of the individuals $i_{s}$. }
    }
    \label{fig:data}
\end{figure}
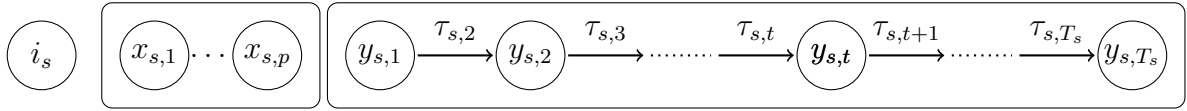

\begin{table}[!ht]
\centering
\begin{tabularx}{\textwidth}{c *{2}{X}}
\hline$\X_{1}$ &  Set of genotypes, $\{F,W\}$ \\\hline
$\X_{2}$ &  Set of social contexts, $\{U,L,A\}$ \\\hline
$\Y$ & Set of four sound syllables, $\{\texttt{d},\texttt{m},\texttt{s},\texttt{u}\}$ \\\hline
$y_{s,t}$ & Sound syllable at time $t$ for sequence $s$, $y_{s,t} \in \Y$ \\\hline
$\tau_{s,t}$ & ISI between {$(t-1)\th$ and $t\th$} syllable of sequence $s$ \\\hline
$x_{s,1}$ & Covariate 1, genotype, $x_{s,1}\in \X_{1}$  \\\hline
$x_{s,2}$ & Covariate 2, social context, $x_{s,2}\in\X_{2}$  \\\hline
{$(y_{s,t-1}, y_{s,t})$} & {Covariate 3, the preceding and the current syllable, $y_{s,t-1} \in \Y,  y_{s,t} \in\Y$}  \\\hline
\end{tabularx}
\vskip 5pt
\begin{tabularx}{\textwidth}{c *{2}{X}}
\multicolumn{2}{c}{{\bf Notations for the Transition Dynamics of the Syllables}} \vspace*{3pt}\\ \cline{1-2}
$z_{trans,j,x_{s,j}}$ & Cluster label of $x_{s,j}$ \\\hline
$\C_{trans}^{(j)}$ & Partition of $\X_{j}$ \\\hline
$k_{trans,j}$ & Number of clusters of partition $\C_{trans}^{(j)}$ \\\hline
{$\blambda_{trans,x_{1},x_{2}}$} & {Transition probability vector for covariate $x_{1}$ and $x_{2}$} \\\hline
$\blambda_{trans,h_{1},h_{2}}$ & Transition probability vector for cluster $h_{1}$ and $h_{2}$ \\\hline
$\blambda^{(i)}_{trans}$ & Transition probability vector for mouse $i$ \\\hline
$\bpi_{trans,0}^{(i)}$ & State-specific probability of fixed effect of mouse $i$ \\\hline
$\bpi_{trans,1}^{(i)}$ & State-specific probability of random effect of mouse $i$ \\\hline
\end{tabularx}
\vskip 5pt
\begin{tabularx}{\textwidth}{c *{2}{X}}
\multicolumn{2}{c}{{\bf Notations for the Distributions of the Inter-Syllable Intervals}}  \vspace*{3pt}\\ \cline{1-2}
$z_{isi,s,t}$ & {Component label of gamma mixtures} \\\hline 
$z_{isi,r,x_{s,r}}$ & Cluster label of $x_{s,r}$ \\\hline
{$z_{isi,r,(y_{s,t-1},y_{s,t})}$} & {Cluster label of $(y_{s,t-1}, y_{s,t})$} \\\hline
$\C_{isi}^{(r)}$ & Partition of $\X_{r}/\Y$ \\\hline
$k_{isi,r}$ & Number of clusters of partition $\C_{isi}^{(r)}$ \\\hline
{$\blambda_{isi,x_{1},x_{2},(y_{t-1},y_t)}$} & {Mixture probability vector for covariate $x_{1}$, $x_{2}$ and syllables $(y_{t-1}, y_t)$} \\\hline
$\blambda_{isi,g_{1},g_{2},g_{3}}$ & Mixture probability vector for cluster $g_{1}$, $g_{2}$ and $g_{3}$ \\\hline
$\blambda^{(i)}_{isi}$ & Mixture probability vector for mouse $i$ \\\hline
$\bpi_{isi,0}^{(i)}$ & Component-specific probability of fixed effect of mouse $i$ \\\hline
$\bpi_{isi,1}^{(i)}$ & Component-specific probability of random effect of mouse $i$\\\hline
\end{tabularx}
\caption{Notations used in our Markov renewal mixed model for vocalization syntax.}
\label{tab: notation}
\end{table}

{We recall that for each sequence $s$,  $(\by,\btau)$ is a Markov renewal process when we have 
\vspace{-6ex}\\
\bse
&& p(y_{s,t}, \tau_{s,t} \mid y_{s,t-1},\tau_{s,t-1},\ldots,y_{s,1},\tau_{s,1}, \btheta_{trans}, \btheta_{isi}) = p(y_{s,t}, \tau_{s,t} \mid y_{s,t-1}, \btheta_{trans}, \btheta_{isi}), 
\ese
\vspace{-6ex}\\
where $\btheta_{trans}$ and $\btheta_{isi}$ are the parameters specifying the transition and ISI distributions, respectively. 
We further assume that (i) the ISI lengths  $\tau_{s,t}$'s depend on both the preceding syllable $y_{s,t-1}$ and the current syllable $y_{s,t}$ (Figure \ref{fig:data}), 
and (ii) the state transitions and the ISI densities do not share any parameters, i.e., $\btheta_{trans} \cap \btheta_{isi} = \phi$. 
Then, we can further write 
\vspace{-6ex}\\
\bse
&& p(y_{s,t}, \tau_{s,t} \mid y_{s,t-1}, \btheta_{trans}, \btheta_{isi}) = p(y_{s,t} \mid y_{s,t-1}, \btheta_{trans}) \cdot p(\tau_{s,t} \mid y_{s,t-1}, y_{s,t}, \btheta_{isi}). 
\ese
\vspace{-6ex}\\}
The joint posterior of $(\btheta_{trans}, \btheta_{isi})$ then also factorizes as 
\vspace{-6ex}\\
\bse
&& p(\btheta_{trans}, \btheta_{isi} \mid \by,\btau) \propto p(\by \mid \btheta_{trans}) p(\btau \mid \btheta_{isi},\by)  p_{0}(\btheta_{trans}) p_{0}(\btheta_{isi}) \\
&&~~\propto p(\btheta_{trans} \mid \by) p(\btheta_{isi} \mid \by,\btau),
\ese
\vspace{-6ex}\\
where $p_{0}(\cdot)$ denotes a generic for the priors. 
In what follows, we thus discuss the models for the syllables and the ISIs separately.

\subsection{Model for Syllable Transitions}

We use the Bayesian mixed effects Markov model of \cite{sarkar2018bayesian} for syllable transitions. 
We describe the model here to keep this article relatively self-contained.

We begin with specifying the transition dynamics as a mixed Markov model as 
\be
\begin{aligned}\label{eq: mixed Markov model}
& \Pr(y_{s,t}=y_{t} \mid  i_{s}=i, x_{s,1}=x_{1}, x_{s,2}=x_{2}, y_{s,t-1}=y_{t-1})=P_{trans,x_{1},x_{2}}^{(i)}(y_{t} \mid y_{t-1}),\\ 
& P_{trans,x_{1},x_{2}}^{(i)} (y_{t} \mid y_{t-1}) = \pi_{trans,0}^{(i)}(y_{t-1}) \lambda_{trans,x_{1},x_{2}}(y_{t} \mid y_{t-1}) + \pi_{trans,1}^{(i)}(y_{t-1}) \lambda^{(i)}_{trans}(y_{t} \mid y_{t-1}). 
\end{aligned}
\ee
The mixed effects transition probabilities $P_{trans,x_{1},x_{2}}^{(i)}(y_{t} \mid y_{t-1})$'s are modeled here as 
a flexible convex mixture of 
a baseline fixed effect component $\blambda_{trans,x_{1},x_{2}}(\cdot\mid y_{t-1})$ for the exogenous covariates, namely genotype and context, 
and a random effect component $\blambda^{(i)}_{trans}(\cdot\mid y_{t-1})$ for the mouse. 
The weights $\bpi_{trans,0}^{(i)}$ and $\bpi_{trans,1}^{(i)}=1-\bpi_{trans,0}^{(i)}$ of the two effects are also allowed to be mouse-specific. 
{In \cite{sarkar2018bayesian}, the coefficient for the convex combination 
did not vary between the individuals but was fixed at $\pi_{trans,1}(y_{t-1})$. 
The population-level model obtained after integrating out the $\lambda^{(i)}_{trans}(y_{t} \mid y_{t-1})$'s was shown to be able to characterize all possible cases of predictor dependent transition probabilities, 
including accommodating all order interactions between them, 
and the individual-level model could also accommodate deviations from it very flexibly. 
Our specification here retains the nonparametric nature of the population-level model 
but by allowing the coefficients $\pi_{trans,1}^{(i)}(y_{t-1})$ to be mouse-specific, 
which accommodates more flexibility in characterizing individual heterogeneity. 
}

For the fixed effects of each covariate $j$, we try to further identify its levels that have a similar effect on the song dynamics. 
This is done by creating a probabilistic partition $\C_{trans}^{(j)}=\{\C_{trans,h_{j}}^{(j)}\}_{h_{j}=1}^{k_{trans,j}}$ of its levels. 
Given partitions $\C_{trans}^{(1)}$ and $\C_{trans}^{(2)}$, 
songs with covariates in the same clusters, say $\C_{trans,h_{1}}^{(1)}$ and $\C_{trans,h_{2}}^{(2)}$, 
then share the same baseline transition probability $\blambda_{trans,h_{1},h_{2}}(\cdot\mid y_{t-1})$.
The specification of the probabilistic partition models is facilitated
by introducing latent cluster allocation variables $\{z_{trans,j,\ell}\}_{j=1,\ell=1}^{2,d_{j}}$, 
with $z_{trans,j,\ell}$ indicating the cluster label for the $\ell\th$ level of the $j\th$ covariate.
Two levels $\ell_{1},\ell_{2} \in \X_{j}=\{1,\dots,d_{j}\}$ will be clustered together if and only if $z_{trans,j,\ell_{1}}=z_{trans,j,\ell_{2}}$. 
For example, $z_{trans,j=2,\ell=2}=z_{trans,j=2,\ell=3}=1$ means that songs produced under contexts 
$L$ $(j=2,\ell=2)$ and $A$ $(j=2,\ell=3)$ 
belong to cluster $\C_{trans,h_{2}=1}^{(j=2)}$ of $\C_{trans}^{(j=2)}$, etc. 
Importantly, when the levels of a covariate $j$ are all clustered together, 
i.e., $k_{trans,j}=1$, 
the transition probabilities do not vary with the levels of covariate $j$. 
The covariate $j$ thus has no effect on the transition dynamics when $k_{trans,j}=1$, 
allowing us to easily and formally test its significance based on the posterior probability of the event $k_{trans,j}=1$.

Keeping these conditioning variables implicit, 
the final transition probability of syllables 
in a song $s$ produced by a mouse $i_{s}=i$ with genotype $x_{s,1}$ in cluster $h_{1}$ and context $x_{s,2}$ in cluster $h_{2}$ is given by 
\bse
\bP_{trans,h_{1},h_{2}}^{(i)} (\cdot\mid y_{t-1}) = \pi_{trans,0}^{(i)}(y_{t-1}) \blambda_{trans,h_{1},h_{2}}(\cdot\mid y_{t-1}) + \pi_{trans,1}^{(i)}(y_{t-1}) \blambda^{(i)}_{trans}(\cdot\mid y_{t-1}).
\ese

We assign conditionally conjugate Dirichlet priors to the fixed effect components
\bse
\blambda_{trans,h_{1},h_{2}}(\cdot\mid y_{t-1}) \sim \Dir\left\{\alpha_{trans,0}\lambda_{trans,0}(1 \mid y_{t-1}),\dots,\alpha_{trans,0}\lambda_{trans,0}(4 \mid y_{t-1})\right\}.
\ese
For the random effect distribution,  for any $y_{t-1} \in \Y$,  we let 
\bse
 \blambda^{(i)}_{trans}(\cdot \mid y_{t-1}) \sim \Dir\left\{\alpha^{(0)}_{trans}\lambda_{trans,0}(1 \mid y_{t-1}),\dots,\alpha^{(0)}_{trans}\lambda_{trans,0}(4 \mid y_{t-1})\right\}.
\ese
We assign a Beta prior to the mouse-specific coefficient for any $y_{t-1} \in \Y$ as  
\bse
\pi_{trans,0}^{(i)}(y_{t-1}) \sim \Beta(a_{trans,0},a_{trans,1}).
\ese
Centering both $\blambda_{trans,h_{1},h_{2}}$ and $\blambda^{(i)}_{trans}$ around the same mean vector $\blambda_{trans,0}$ facilitates posterior computation.
The random effects $\blambda^{(i)}_{trans}$ and $\bpi^{(i)}_{trans,0}$ can be easily integrated out 
to obtain a closed-form expression for population-level probabilities as
\bse
\bP_{trans,h_{1},h_{2}} (\cdot\mid y_{t-1}) = \pi_{trans,0}\blambda_{trans,h_{1},h_{2}}(\cdot\mid y_{t-1}) + \pi_{trans,1}\blambda_{trans,0}(\cdot\mid y_{t-1}), \label{eq:trans_pop}
\ese
where 
\bse
\pi_{trans,0} = \frac{a_{trans,0}}{a_{trans,0}+a_{trans,1}}, ~~~\pi_{trans,1} = \frac{a_{trans,1}}{a_{trans,0}+a_{trans,1}}.
\ese
We assume that some states in $\Y$ are naturally preferred regardless of the values of the covariates. 
To capture this, we let $\blambda_{trans,0}$ center around a global $\blambda_{trans,00}$. 
Lastly, the hyper-parameters $\alpha^{(0)}_{trans}$ and $\alpha_{trans,0}$ are given gamma hyper-priors.

The complete Bayesian hierarchical model for the transition dynamics can then be summarized as follows. 
\vspace{-4ex}\\
\be
\begin{aligned}\label{eq: MEMSEP bayesian}
& (y_{s,t} \mid y_{s,t-1}=y_{t-1}, i_{s}=i, z_{trans,1,x_{s,1}} =h_{1}, z_{trans,2,x_{s,2}}=h_{2})  \sim \\
& \hspace{3cm} \Mult\left\{P_{trans,h_{1},h_{2}}^{(i)}(1\mid y_{t-1}),\dots,P_{trans,h_{1},h_{2}}^{(i)}(4\mid y_{t-1})\right\},  \\
& \bP_{trans,h_{1},h_{2}}^{(i)} (\cdot\mid y_{t-1}) = \pi_{trans,0}^{(i)}(y_{t-1}) \blambda_{trans,h_{1},h_{2}}(\cdot\mid y_{t-1}) + \pi_{trans,1}^{(i)}(y_{t-1}) \blambda^{(i)}_{trans}(\cdot\mid y_{t-1}),\\
& z_{trans,j,\ell} \sim \Mult\left\{\mu_{trans,j}(1),\dots,\mu_{trans,j}(d_{j})\right\}, ~~~ \bmu_{trans,j} \sim \Dir(\alpha_{trans,j},\dots,\alpha_{trans,j}), \\
& \blambda_{trans}^{(i)}(\cdot \mid y_{t-1}) \sim \Dir\left\{\alpha^{(0)}_{trans}\lambda_{trans,0}(1 \mid y_{t-1}),\dots,\alpha^{(0)}_{trans}\lambda_{trans,0}(4 \mid y_{t-1})\right\}, \\
& \blambda_{trans,h_{1},h_{2}}(\cdot\mid y_{t-1}) \sim \Dir\left\{\alpha_{trans,0}\lambda_{trans,0}(1 \mid y_{t-1}),\dots,\alpha_{trans,0}\lambda_{trans,0}(4 \mid y_{t-1})\right\}, \\
& \blambda_{trans,0}(\cdot\mid y_{t-1}) \sim \Dir\left\{\alpha_{trans,00}\lambda_{trans,00}(1),\dots,\alpha_{trans,00}\lambda_{trans,00}(4)\right\},  \\
& \pi_{trans,0}^{(i)}(y_{t-1}) \sim \Beta(a_{trans,0},a_{trans,1}),  \\
& \alpha_{trans,0}\sim\Ga(a_{trans,0},b_{trans,0}),~~~~~~~\alpha^{(0)}_{trans}\sim\Ga(a^{(0)}_{trans},b^{(0)}_{trans}).
\end{aligned}
\ee

\subsection{Model for Inter-Syllable Intervals}

For the inter-syllable interval (ISI) times $\{\tau_{s,t}\}_{s=1,t=2}^{s_0,T_s}$, 
we associate each ISI from song $s$ with $3$ covariates: 
two exogenous, namely genotype $x_{s,1}\in \X_1=\{1,2\}$ and social context $x_{s,2}\in\X_2=\{1,2,3\}$, 
{and one local, namely the pair of the preceding and the current syllable $(y_{s,t-1}, y_{s,t}) \in \Y \times \Y = \{1,2,3,4\} \times \{1,2,3,4\}$}. 
Given the values of these covariates, 
{we model the log-transformed ISI times $\wt\tau_{s,t} = \log(\tau_{s,t}+1)$ of the mouse $i_s=i$} using mixtures of gamma kernels as 
\vspace{-3ex}\\
\be
\begin{aligned}\label{eq:isi}
& {f(\wt\tau_{s,t} \mid  i_{s}=i, x_{s,1}=x_{1}, x_{s,2}=x_{2}, (y_{s,t-1}, y_{s,t})=(y_{t-1},y_t))} \\
&\hspace{1cm}{= \sum_{k=1}^{K}P_{isi}^{(i)}(k \mid x_{1},x_{2},(y_{t-1}, y_t)) \Ga(\wt\tau_{s,t} \mid \alpha_{k},\beta_{k}), }
\end{aligned}
\ee
\vspace{-3ex}\\
where $\Ga(\cdot\mid\alpha_{k},\beta_{k})$ denotes a gamma mixture kernel with shape $\alpha_{k}$ and rate $\beta_{k}$
and $K$ is the total number of mixture components.
{$P_{isi}^{(i)}(k \mid x_{1},x_{2},{(y_{t-1}, y_t)})$}'s are mixed effects mixture probabilities that vary with the associated covariate values and are also specific to the subject. 
Introducing latent variables $\{z_{isi,s,t}\}$ indicating the index of the mixture component, 
we can write 
\vspace{-3ex}\\
\be
\begin{aligned} \label{eq: mixed mixture model}
& f(\wt\tau_{s,t} \mid  z_{isi,s,t}=k) \sim \Ga(\wt\tau_{s,t} \mid \alpha_{k},\beta_{k}), \\
& {\Pr(z_{isi,s,t}=k \mid  i_{s}=i, x_{s,1}=x_{1}, x_{s,2}=x_{2}, (y_{s,t-1},y_{s,t})=(y_{t-1},y_t))=P_{isi}^{(i)}(k \mid x_{1},x_{2},(y_{t-1}, y_t))}. 
\end{aligned}
\ee
 \vspace{-3ex}\\
Model (\ref{eq: mixed mixture model}) is structurally similar to model (\ref{eq: mixed Markov model}) 
except that we are now modeling the distribution of a \emph{latent} categorical variable $z_{isi,s,t}$ as opposed to the \emph{observed} categorical variable $y_{s,t}$. 
The number of components $K$ in (\ref{eq:isi}) is thus also unknown and needs to be inferred from the data, bringing in significant additional challenges. 
Nevertheless, we can use similar strategies to model the mixed effects mixture probabilities {$P_{isi}^{(i)}(k \mid x_{1},x_{2},(y_{t-1}, y_t))$} as  
\bse
&& {\bP^{(i)}_{isi,x_{1},x_{2},{(y_{t-1}, y_t)}}(\cdot)=\pi_{isi,0}^{(i)}(\cdot)\blambda_{isi,x_{1},x_{2},{(y_{t-1}, y_t)}}(\cdot)+\pi_{isi,1}^{(i)}(\cdot)\blambda^{(i)}_{isi}(\cdot),}
\ese
where {$\blambda_{isi,x_{1},x_{2},{(y_{t-1}, y_t)}}(\cdot)$} is the fixed effect component for the associated covariates, namely genotype, context and the {preceding-current syllable pair}, 
and $\blambda^{(i)}_{isi}(\cdot)$ is the random effect component for the mouse. 
The weights $\pi_{isi,0}^{(i)}$ and $\pi_{isi,1}^{(i)}=1-\pi_{isi,0}^{(i)}$ of the two effects are also allowed to be mouse-specific, as before.

To assess the significance of each covariate $r$, 
we induce a clustering $\C_{isi}^{(r)}=\{\C_{isi,g_{r}}^{(r)}\}_{g_{r}=1}^{k_{isi,r}}$ of its levels 
so that 
ISIs with associated covariates in the same clusters, say $g_{1}$, $g_{2}$ and $g_{3}$, share the same fixed effect mixture probability component $\blambda_{isi,g_{1},g_{2},g_{3}}$. 
This is done via introducing latent cluster allocation variables $\{z_{isi,r,w}\}_{r=1,w=1}^{3,d_{r}}$, as before, 
with $z_{isi,r,w}$ indicating the cluster label for the $w\th$ level of the $r\th$ covariate.
Importantly, as in the case of transition probabilities, 
when the levels of a covariate $r$ are all clustered together, 
i.e., $k_{isi,r}=1$, 
the covariate $r$ has no effect on the ISI distribution, 
allowing us to easily and formally test its significance based on the posterior probability of the event $k_{isi,r}=1$. 

Keeping these conditioning variables implicit, 
the ISI mixture probability
in a song $s$ produced by a mouse $i_{s}=i$ with genotype $x_{s,1}$ in cluster $g_{1}$ and context $x_{s,2}$ in cluster $g_{2}$ and {syllable pair $(y_{s,t-1}, y_{s,t})$ in cluster $g_{3}$} 
is given by 
\bse
\bP_{isi,g_{1},g_{2},g_{3}}^{(i)} (\cdot) = \pi_{isi,0}^{(i)}(\cdot) \blambda_{isi,g_{1},g_{2},g_{3}}(\cdot) + \pi_{isi,1}^{(i)}(\cdot) \blambda^{(i)}_{isi}(\cdot).
\ese

As earlier, we assign conditionally conjugate Dirichlet priors to the fixed and random effect components, and give a Beta prior to the mouse-specific coefficient,  
\bse
\begin{aligned}
& \blambda_{isi,g_{1},g_{2},g_{3}}(\cdot) \sim \Dir\left\{\alpha_{isi,0}\lambda_{isi,0}(1),\dots,\alpha_{isi,0}\lambda_{isi,0}(K)\right\}, \\
& \blambda^{(i)}_{isi}(\cdot) \sim \Dir\left\{\alpha^{(0)}_{isi}\lambda_{isi,0}(1),\dots,\alpha^{(0)}_{isi}\lambda_{isi,0}(K)\right\}, \\
& \pi_{isi,0}^{(i)}(k) \sim \Beta(a_{isi,0},a_{isi,1}).
\end{aligned}
\ese

The random effects $\blambda^{(i)}_{isi}$ and $\pi^{(i)}_{isi,0}$ can be easily integrated out 
to obtain the closed-form population-level mixture probabilities as
\bse
P_{isi,g_{1},g_{2},g_{3}} (k) = \pi_{isi,0}\lambda_{isi,g_{1},g_{2},g_{3}}(k) + \pi_{isi,1}\lambda_{isi,0}(k), \label{eq:isi_pop}
\ese
where $\pi_{isi,0} = \frac{a_{isi,0}}{a_{isi,0}+a_{isi,1}}$ and $\pi_{isi,1} = \frac{a_{isi,1}}{a_{isi,0}+a_{isi,1}}$. 
Finally, we let $\blambda_{isi,0}$ center around a global $\blambda_{isi,00}$ 
and the hyper-parameters $\alpha^{(0)}_{isi}$ and $\alpha_{isi,0}$ are given gamma hyper-priors.

The complete Bayesian hierarchical model for the ISIs may be summarized as 
\be
\begin{aligned}\label{eq: isi bayesian}
& (\wt\tau_{s,t} \mid z_{isi,s,t}=k) \sim \Ga(\wt\tau_{s,t} \mid \alpha_{k},\beta_{k}), \\
& (z_{isi,s,t} \mid i_{s}=i, z_{isi,1,x_{s,1}} =g_{1}, z_{isi,2,x_{s,2}}=g_{2}, {z_{isi,3,{(y_{s,t-1}, y_{s,t})}}=g_{3}})  \sim\\
& \hspace{3cm} \Mult\left\{P_{isi,g_{1},g_{2},g_{3}}^{(i)}(1),\dots,P_{isi,g_{1},g_{2},g_{3}}^{(i)}(K)\right\}, \\
& P^{(i)}_{isi,g_{1},g_{2},g_{3}}(k)=\pi_{isi,0}^{(i)}(k)\lambda_{isi,g_{1},g_{2},g_{3}}(k)+\pi_{isi,1}^{(i)}(k)\lambda^{(i)}_{isi}(k) \\
& \blambda_{isi}^{(i)}(\cdot) \sim \Dir\left\{\alpha_{isi}^{(0)}\lambda_{isi,0}(1),\dots,\alpha_{isi}^{(0)}\lambda_{isi,0}(K)\right\}, ~~~\alpha_{isi}^{(0)}\sim\Ga(a_{isi}^{(0)},b_{isi}^{(0)}), \\
& \blambda_{isi,g_{1},g_{2},g_{3}}(\cdot) \sim \Dir\left\{\alpha_{isi,0}\lambda_{isi,0}(1),\dots,\alpha_{isi,0}\lambda_{isi,0}(K)\right\}, ~~~\alpha_{isi,0}\sim\Ga(a_{isi,0},b_{isi,0}), \\
&\blambda_{isi,0}(\cdot) \sim \Dir\left\{\alpha_{isi,00}\lambda_{isi,00}(1),\dots,\alpha_{isi,00}\lambda_{isi,00}(K)\right\},\\
& \pi_{isi,0}^{(i)}(k) \sim \Beta(a_{isi,0},a_{isi,1}),\\ 
& \alpha_{k} \sim \Ga(a_{isi,0},b_{isi,0}), ~~~\beta_{k} \sim \Ga(a_{isi,0},b_{isi,0}).
\end{aligned}
\ee

Gamma mixtures, in other forms, have appeared before in \cite{chen2000probability, wiper2001mixtures, hanson2006modeling}, etc. 
{Such mixtures can approximate a large class of distributions on $[0,\infty)$ \citep[see, e.g., Theorem 14 in][]{wu2008kullback}.} 
The gamma kernel, however, brings in computational challenges which we address briefly in Section \ref{sec: hyper-parameters post inf} below 
and then in detail again in {Appendix D} in the supplementary materials. 
There also exists some literature on flexible mixture and partition models for conditionally varying densities of continuous random variables in the presence of covariates  \citep[etc.]{maceachern1999dependent, chung_dunson:2009, muller2011product}. 
To our knowledge, however, flexible mixed effects mixture models of the type proposed here that accommodate both fixed effects of covariates as well as random heterogeneity of subjects while also allowing simultaneous covariate selection have not appeared in the literature before. 
\section{Prior Hyper-parameters and Posterior Inference} \label{sec: hyper-parameters post inf}

The choice of prior hyper-parameters, 
including the choice of the number of mixture components $K$ in the gamma mixture model (\ref{eq:isi}) for which we use predictive model selection criteria, 
are discussed in {Appendix B} in the supplementary materials.

Our posterior inference is based on samples of the model parameters drawn using an  MCMC algorithm.
The full conditional posterior distributions are mostly obtained in closed-form and hence are easy to sample from. 
One exception is the sampling of the gamma mixture parameters.
The conjugate prior for gamma distribution with unknown shape parameter is known to be analytically intractable \citep{damsleth1975conjugate, miller1980bayesian}, 
posing difficulty in sampling the parameters $\alpha_{k}$'s and $\beta_{k}$'s.
We experimented a number of sampling methods and ultimately used the strategy introduced in \cite{miller2019fast}
which uses a gamma density function to approximate the full conditional for gamma shape parameters. 
After sampling the shape parameters $\alpha_{k}$'s, the rate parameters $\beta_{k}$'s can be easily sampled from their closed-form conjugate gamma full conditionals.
As the results of Sections \ref{sec: real data results} and {Appendix~F} will illustrate, this method worked well with real data as well as in our simulation experiments, 
converging quickly, mixing well and providing accurate estimates of the target distributions.
Details of the posterior sampling algorithm can be found in {Appendix~D} in the supplementary materials.

\section{Results for the Foxp2 Data Set} \label{sec: real data results}

In this section, we discuss the results of the proposed Bayesian Markov renewal mixed model fitted to the Foxp2 data set.  
We present the results for syllable transitions and ISIs separately. 
In Section \ref{sec:res-trans}, we analyze the influence of genotypes and social contexts on syllable transitions. 
{In Section \ref{sec:res-isi}, we analyze the influence of genotypes, social contexts and the {preceding-current syllable pairs} on the distribution of the ISIs.}

\subsection{Results for Syllable Transition}\label{sec:res-trans}

\begin{figure}[ht!]
\centering
\includegraphics[width=\textwidth]{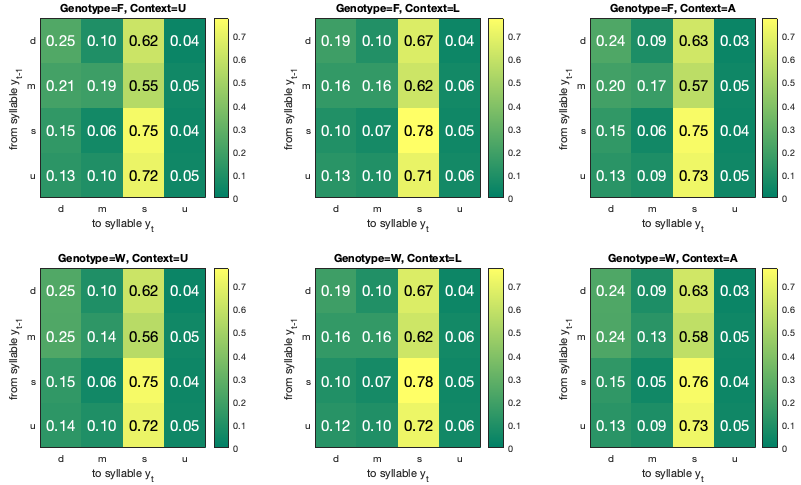}
\caption{Results for transitions for the Foxp2 data set showing the estimated posterior mean of the transition probabilities $P_{trans,x_1,x_2}(y_t \mid y_{t-1})$ of syllables $y_t, y_{t-1} \in \Y = \{\texttt{d},\texttt{m},\texttt{s},\texttt{u}\}$ under different combinations of genotype $x_1\in \{F,W\}$ and social context $x_2\in\{U,L,A\}$.}
\label{fig: trans_mean}
\end{figure}

Figure \ref{fig: trans_mean} shows the estimated posterior mean of the population-level transition probabilities,  $P_{trans,x_1,x_2}(y_t \mid y_{t-1})$, given genotype $x_1$ and social context $x_2$. 
We see that regardless of the covariate values, the \texttt{s} syllable is predominantly transitioned to and \texttt{u} is the least likely syllable to transition to across different genotypes and contexts. 
This is reasonable since the \texttt{s} syllable is presumably the easiest to pronounce and \texttt{u} is the least pronounced syllable across all sequences.

\begin{figure}[ht!]
\centering
\includegraphics[width=0.3\textwidth]{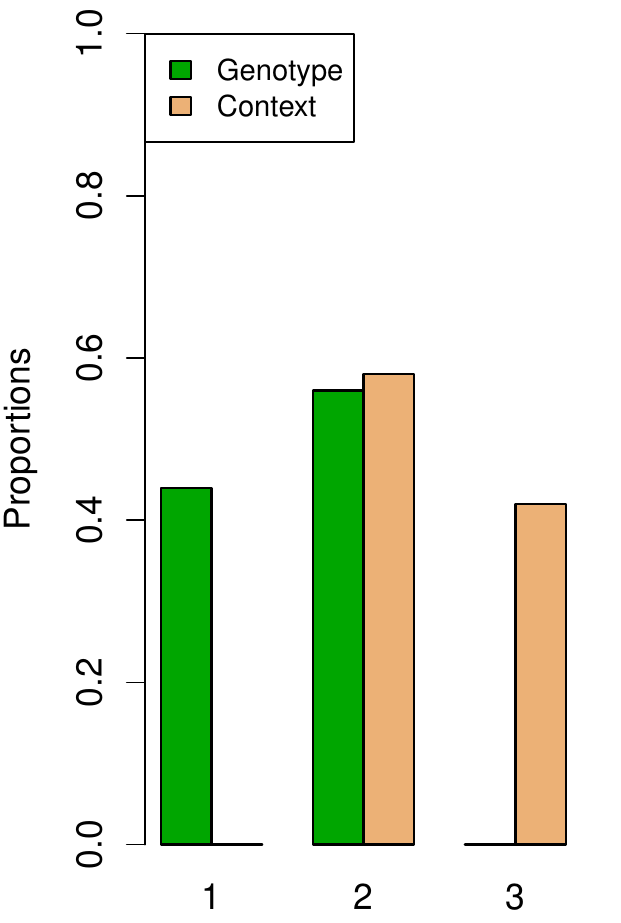}
\caption{Results for the Foxp2 data set showing the proportions of the number of clusters $k_{trans,j}$ for each covariate $j$ among thinned samples after burn-ins for syllable transitions.
The x-axis represents the number of clusters. 
The green/orange bars represent the posterior distribution of $P(k_{trans,j})$ for genotype $x_{j=1} \in \{F,W\}$ and social context $x_{j=2} \in \{U,L,A\}$, respectively.}
\label{fig: trans_global}
\end{figure}

Consider the following hypotheses for the impact of genotype and social context.
\begin{itemize}
\setlength\itemsep{.3em}
\item $H_{trans,0,j}$: The population-level transition probabilities $P_{trans,x_1,x_2}(y_t \mid y_{t-1})$ do not vary based on values of $x_{j}, j=1,2$. 
\item $H_{trans,1,j}$: The population-level transition probabilities $P_{trans,x_1,x_2}(y_t \mid y_{t-1})$ vary based on values of $x_{j}, j=1,2$. 
\end{itemize}
By the construction of our model, $H_{trans,1,j}$ is true if and only if $k_{trans,j}>1$.
Figure~\ref{fig: trans_global} displays the estimated posterior distribution of $k_{trans,j}$ for genotype $(j=1)$ and social context $(j=2)$.
The estimated posterior probability that $k_{trans,1}$ greater than $1$ is $\wh{P}(k_{trans,1}>1 \mid \data)\approx 0.56$.
In contrast to the findings reported previously in \cite{sarkar2018bayesian}, 
the evidence that the mutation on the Foxp2 gene impacts the transition probabilities of the syllables has thus become somewhat weaker.  
This illustrates that treating ISI as a separate continuous variable results in less distinctions between the transition dynamics of the two groups.
Conforming to the previous analyses in \cite{sarkar2018bayesian}, 
there is, however, very clear evidence of the influence of social context on the transitions probabilities. 
In Figure~\ref{fig: trans_global}, we see that the estimated posterior probability $\wh{P}(k_{trans,2}>1 \mid \data)=1$, 
with $\wh{P}(k_{trans,2}=2 \mid \data)=0.58$ and $\wh{P}(k_{trans,2}=3 \mid \data)=0.42$.
Specifically, whenever there were two clusters, the contexts $U$ and $A$ were clustered together. 
In Figure~\ref{fig: trans_mean}, we see that there is strong evidence that contexts $U$ and $A$ have similar impact across genotypes. 
Compared to $U$ and $A$, the $L$ context has smaller transition probabilities for transitions types $\texttt{d} \rightarrow \texttt{d}$ and $\texttt{m} \rightarrow \texttt{d}$ across the two genotypes.
The decrease in the transition probabilities of $\texttt{d} \rightarrow \texttt{d}$ and $\texttt{m} \rightarrow \texttt{d}$ 
seems to be explained by the increase in the probabilities of $\texttt{d} \rightarrow \texttt{s}$ and $\texttt{m} \rightarrow \texttt{s}$, 
suggesting that the mice vocalized short and simple symbols more often under the $L$ context.

{In \cite{sarkar2018bayesian}, the coefficients $(\bpi_{trans,0},\bpi_{trans,1})$ were assumed to be shared between all mice, whereas here we have allowed them to be mouse-specific as $(\bpi_{trans,0}^{(i)},\bpi_{trans,1}^{(i)})$. 
To investigate the effectiveness of this, 
we looked at the values for these coefficients for different mice. 
Given a preceding syllable $y_{t-1}$, we compare the coefficient $\bpi_{trans,0}^{(i)}$ for each individual $i$.
We present these results in Table~3 in Appendix E in the Supplementary Materials.
From the table, we see that these coefficients differ substantially between different mice, especially for preceding syllables \texttt{u} and \texttt{m}, justifying our decision of making them mouse-specific.}

\subsection{Results for Inter-Syllable Intervals}\label{sec:res-isi}
We first transform the observed ISIs to $\wt\tau_{s,t}=\log(1+\tau_{s,t})$. 
The original $\tau_{s,t}$'s have a wide range with a minimum of $0.01$ and a maximum of $258.8$ seconds. 
This pre-processing step helps shorten the range of the data and produce better graphical summaries for the results.
We add $1$ to the original $\tau_{s,t}$'s before taking the log to avoid negative $\wt\tau_{s,t}$'s.

\begin{figure}[!ht]
\centering
\subfloat[Histogram of the transformed ISIs with the estimated posterior mean (red line) of their  marginal gamma mixture density based on MCMC samples after burn-in and thinning.]{
\includegraphics[width=0.43\textwidth]{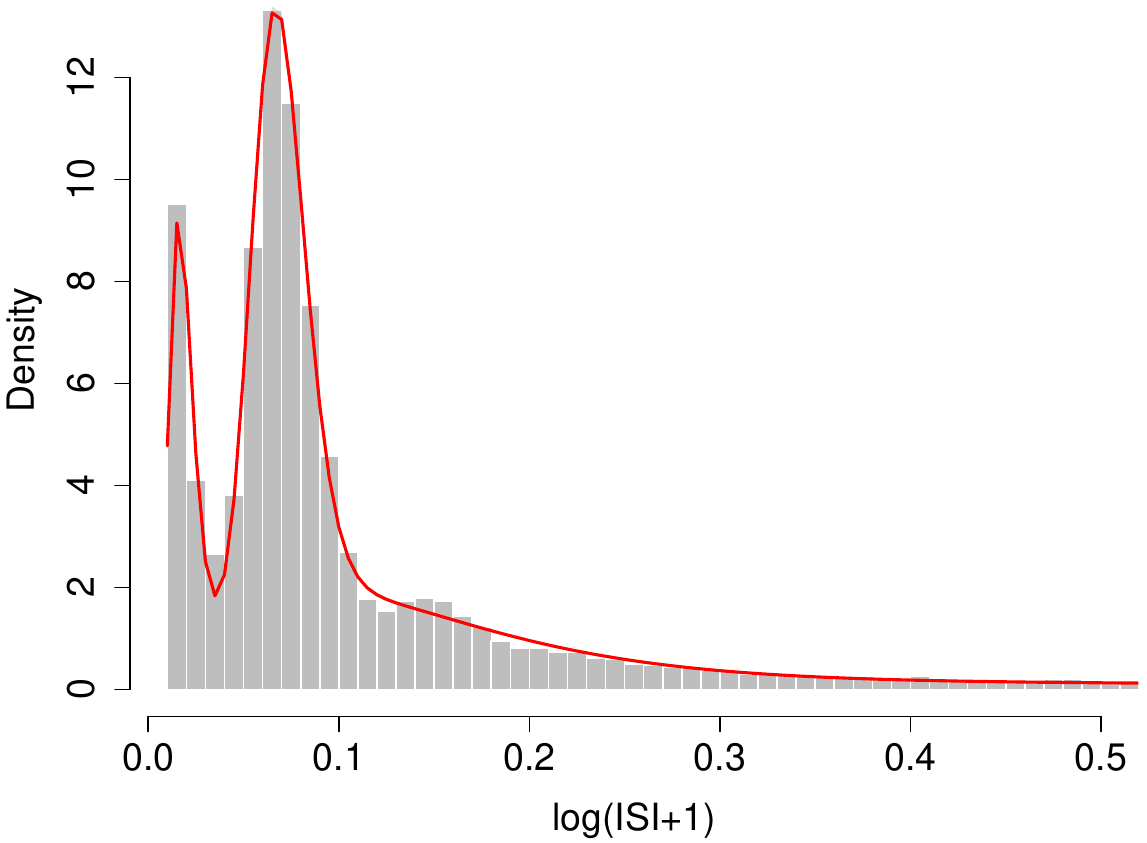}
\label{fig:isi_hist_total}}
\qquad
\subfloat[Histograms of the transformed ISIs for each component of the gamma mixture model along with the component density (red lines) from the last MCMC iteration. The x-axes are adjusted for better visualization.]{
\includegraphics[width=0.47\textwidth]{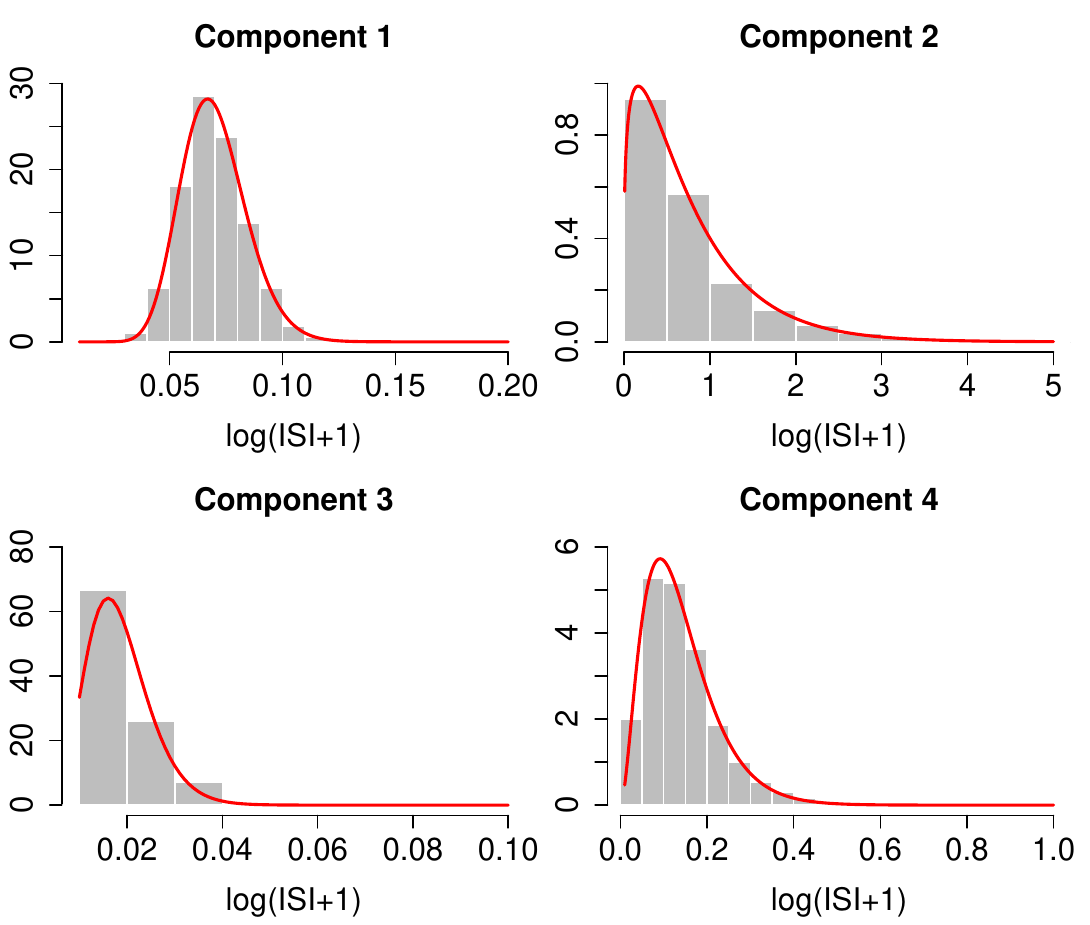}
\label{fig:isi_hist_comp}}\\
\caption{Results for ISIs for the Foxp2 data set.}
\label{fig:isi_hist_post}
\end{figure}

Figure \ref{fig:isi_hist_total} shows the histogram of the transformed ISIs along with the posterior mean (red curve) averaged from samples after burn-ins and thinning. 
{Table~\ref{tab: isi mean} in Section \ref{sec: data set and background} shows the empirical and posterior means and standard deviations of the ISIs side-by-side}. 
We see that our proposed model fits the ISI data very well, whose two peaks are captured by the mixture gamma distribution. 
Figure \ref{fig:isi_hist_comp} displays the histogram for each mixture component 
along with the corresponding gamma densities with shape and rate parameters taken from the last iteration of the MCMC sampler.
{It is clear that components {1 and 3} represent the two peaks we see in Figure \ref{fig:isi_hist_total} 
whereas components {2} and 4 assign more probability mass to larger values of the transformed ISIs.} 
The shape and rate parameters for the gamma distribution of each component are presented in {the first table in Table~\ref{tab:isi_heatmap}}.
The shape and rate parameters for components 1 and {3} are much larger than those of components {2} and 4, 
capturing the concentration of the small values of the transformed ISIs.

\begin{table}[ht]
\parbox{.29\linewidth}{
\centering
\begin{tabular}{ccc}
&shape.k & rate.k  \\ \cline{2-3}
Comp 1 & $23.47$                   & $336.41$         	 \\
Comp 2 & $1.28$                     & $1.69$          	\\
Comp 3 & $7.79$                 & $423.98$           \\
Comp 4 &$2.89$                   & $20.59$          \\\cline{2-3}
\end{tabular}
\caption*{Estimated gamma shape and rate parameters}
}
~~~
\parbox{.29\linewidth}{
\centering
\begin{tabular}{ccc}
&$F$ & $W$  \\ \cline{2-3}
Comp 1 & $0.52$                   & $0.57$         	 \\
Comp 2 &  $0.16$                     & $0.08$          	\\
Comp 3 &  $0.08$                 & $0.11$           \\
Comp 4 & $0.24$                   & $0.25$          \\\cline{2-3}
\end{tabular}
\caption*{Estimated mixture probabilities for genotypes}
}
~
\parbox{.29\linewidth}{
\centering
\begin{tabular}{cccc}
&$U$ & $L$ & $A$  \\ \cline{2-4}
Comp 1 & $0.56$                   & $0.42$       & $0.65$   	 \\
Comp 2&  $0.15$                     & $0.11$       & $0.09$      	\\
Comp 3&  $0.05$                 & $0.18$         & $0.05$     \\
Comp 4 & $0.23$                   & $0.29$        & $0.21$     \\\cline{2-4}
\end{tabular}
\caption*{Estimated mixture probabilities for contexts}
} 
\newline
\vspace*{5pt}
\newline
\parbox{\linewidth}{
\centering
\begin{tabular}{ccccccccc}
&$(\ttd,\ttd)$ & $(\ttd,\ttm)$ & $(\ttd,\tts)$ & $(\ttd,\ttu)$ &$(\ttm,\ttd)$ & $(\ttm,\ttm)$ & $(\ttm,\tts)$ & $(\ttm,\ttu)$ \\ \cline{2-9}
Comp 1 & 0.65 &0.65 &0.45& 0.47& 0.65& 0.65 &0.47& 0.55   	 \\
Comp 2&  0.07 &0.07& 0.16 &0.12& 0.07& 0.07 &0.12 &0.14      	\\
Comp 3&  0.05 &0.05 &0.10& 0.17& 0.05 &0.05 &0.17 &0.07     \\
Comp 4 & 0.23  &0.23 & 0.29  &0.24  &0.23 & 0.23  &0.24 & 0.24    \\\cline{2-9}
\end{tabular}
\caption*{Estimated mixture probabilities for each preceding-current syllable pair}
}
\newline
\vspace*{5pt}
\newline
\parbox{\linewidth}{
\centering
\begin{tabular}{ccccccccc}
&$(\tts,\ttd)$ & $(\tts,\ttm)$ & $(\tts,\tts)$ & $(\tts,\ttu)$ &$(\ttu,\ttd)$ & $(\ttu,\ttm)$ & $(\ttu,\tts)$ & $(\ttu,\ttu)$\\ \cline{2-9}
Comp 1 & 0.56 & 0.65 & 0.35 & 0.55 & 0.47 &  0.54 & 0.47 & 0.56	 \\
Comp 2&  0.14 & 0.07 & 0.23  &0.14  &0.12  &  0.11  &0.12  &0.14  	\\
Comp 3& 0.06&  0.05  &0.11  &0.07  &0.17 &   0.11 & 0.17 & 0.06 \\
Comp 4 & 0.24 & 0.23&  0.31 & 0.24 & 0.24 &  0.24 & 0.24 & 0.24  \\\cline{2-9}
\end{tabular}
\caption*{Estimated mixture probabilities for each preceding-current syllable pair (cont'd)}
}
\caption{{Results for ISIs for the Foxp2 data set taken from the last MCMC iteration.}}
\label{tab:isi_heatmap}
\end{table}

{Table~\ref{tab:isi_heatmap} displays the mixture probabilities taken from the last MCMC iteration for each covariate: 
genotype ($F$ and $W$), social contexts ($U$, $L$, and $A$), and {every preceding-current syllable pair}. } 
We see that mice with the Foxp2 mutation have a smaller mixture probability in the components 1, {3} and 4 compared to wild-types but have a significantly higher probability for component {2} ($+0.08$). 
Recall that component {2} has the smallest rate parameter, which indicates a higher probability to have a larger value compared to the other components.
A large mixture probability in component {2} indicates a high ISI value. 
This suggests that the ISI length for mice with the Foxp2 mutation concentrates on larger values than the ISIs of wild-types,  
which supports our hypothesis that mice with such mutation needs a longer ISI before pronouncing a new syllable.
One interesting discovery from {Tables~\ref{tab: isi mean} and \ref{tab:isi_heatmap}} is that male mice in the presence of a live female ($L$ context) tend to have shorter ISIs 
than those under the other two contexts since the mixture probability is {much higher in component 3 for the $L$ context compared to the other contexts}.
This suggests that male mice need a shorter interval between pronouncing two syllables in the presence of a live female. 
This finding, along with the discovery that transitions to the simplest syllable \texttt{s} are more frequent under the $L$ context, 
shows that male mice exhibit different vocalization dynamics when there is an awake female mouse present. 

{The last two tables in Table~\ref{tab:isi_heatmap} show the mixture probabilities associated with each preceding-current syllable pair. 
The 16 pairs of syllables have similar weights in the 4 components except for  $(\tts,\tts)$, which has a much larger mixture probability for component 2,   indicating a longer ISI between pronouncing consecutive $\tts$ syllables. 
This corresponds to the result in Table~\ref{tab: isi mean} where the ISIs are usually longer when both the preceding and the current syllable are \texttt{s}, no matter what the genotype or social context is.}

\begin{figure}[ht!]
\centering
\includegraphics[width=0.6\textwidth]{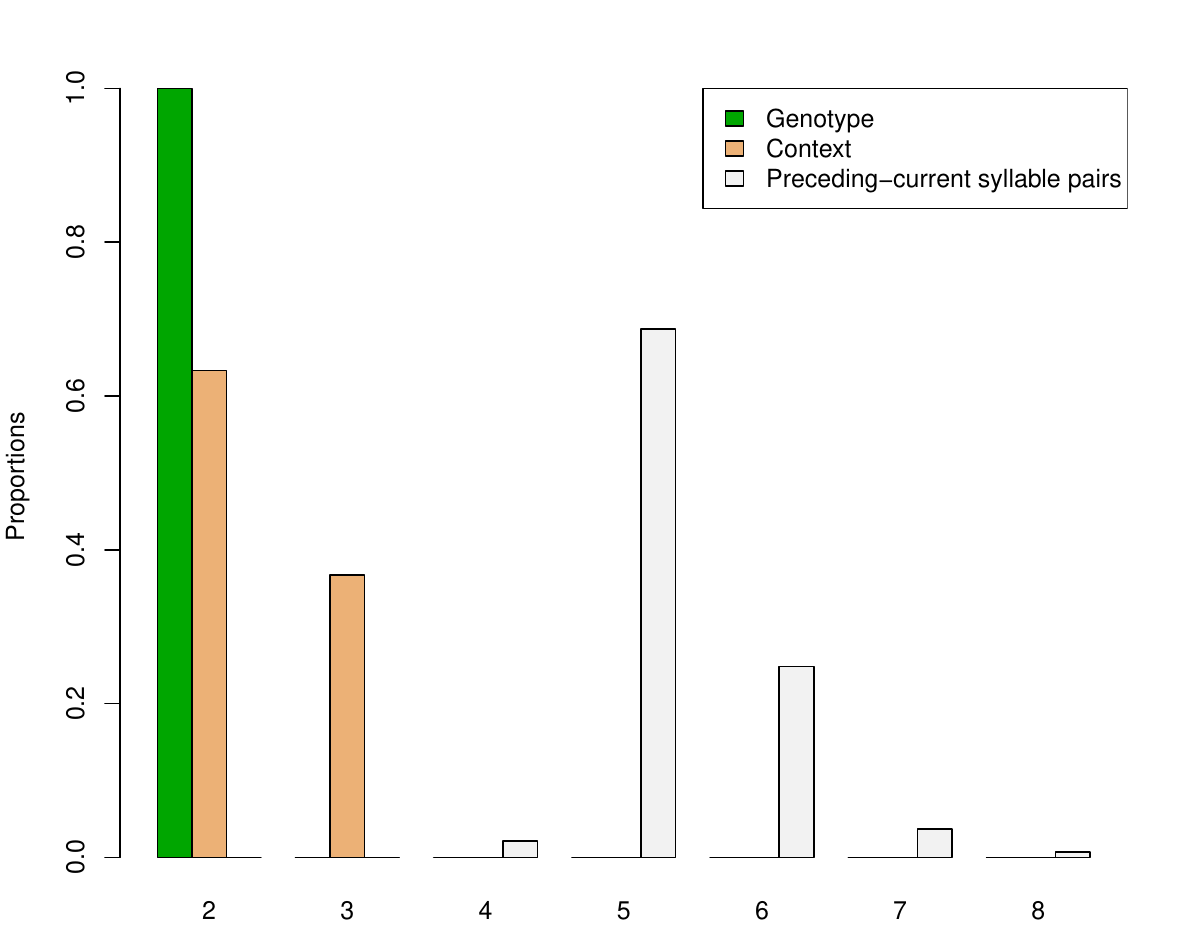}
\caption{Results for the Foxp2 data set showing the proportions of the number of clusters $k_{isi,r}$ for each covariate $r$ among thinned samples after burn-ins.
The x-axis represents the number of clusters. 
The green/orange/white bars represent the posterior distribution of $P(k_{isi,r})$ for genotype $x_1 \in \{F,W\}$, social context $x_2 \in \{U,L,A\}$ and {preceding-current syllable pair $(y_{t-1},y_t) \in \Y \times \Y = \{\texttt{d},\texttt{m},\texttt{s},\texttt{u}\} \times  \{\texttt{d},\texttt{m},\texttt{s},\texttt{u}\}$, respectively.}}
\label{fig: isi_global}
\end{figure}

We are also interested in testing the significance of the impact each covariate has on the ISIs.
In particular, let $H_{isi,0, r}$ be the hypothesis that the distribution of ISI does not vary with the values of the covariate $r$. 
Similar to syllable transitions, $H_{isi,0,r}$ is true if and only if $k_{isi,r}=1$. 
In Figure~\ref{fig: isi_global}, we plot 
the estimated posterior probabilities $\wh{P}(k_{isi,r} \mid \data)$ for each covariate $r$.
{We have $\wh{P}(k_{isi,r} {> 1} \mid \data)=1$ for {$r=1,2, 3$}, implying strong evidence of the influence of genotypes, social contexts and {preceding-current syllable pairs} on the distribution of the ISIs.
Particularly for genotype, $\wh{P}(k_{isi,r=1} = 2 \mid \data)=1$ indicates that both levels of genotypes are significant for the length of ISIs. 
On the other hand, not every level of the social contexts and the syllable pairs is significant. 
Since the number of clusters for every covariate is at least two, we conclude that all three covariates are  significant for the distribution of ISIs.}

\section{Discussion} \label{sec: discussion}
This paper introduced a new class of Bayesian Markov renewal mixed effects models 
that allows inference of both state transition probabilities and continuous inter-state interval times. 
On the statistical side, our main novel contribution is a mixed effects gamma mixture model for the inter-state intervals. 
The mixture probabilities build on carefully constructed convex combinations 
of a fixed effect component for the associated covariates and a random effect component for the associated individual, 
resulting in a highly flexible and computationally tractable model.
At the same time, covariate values that induce similar effects on the response are probabilistically clustered together and significant covariates are identified. 

We used the model to reanalyze the Foxp2 data set 
which comprises a collection of songs sung by adult male mice with or without a Foxp2 mutation under various social contexts.
In contrast to previous analyses, 
we found weaker evidence that the transition dynamics of the syllables within the songs vary with genotype and social context. 
On the other hand, there is significant evidence that all three covariates, namely genotypes, social contexts, and {preceding-current syllable pairs}, influence the lengths of the ISIs. 
On the application side, the important scientific implication is that the vocal impairment of the Foxp2 mice is manifested in their having longer ISIs than wild-types, 
not just in the syllable transition dynamics as previous analyses suggested.

Our proposed model for the transition dynamics builds on the work of \cite{sarkar2018bayesian} where they discretized the ISIs into (consecutive) `silent' states and then modeled the transitions between the states. 
It is difficult to directly compare our work with theirs as they did not model the continuous ISIs separately and, as a result, the influences of the covariates on the transition probabilities and the ISIs are mixed together. 
By separating the two inferences, we found that the influences of both genotype and social context are weaker for transition probabilities among the original four syllables. 
On the other hand, the covariate effects for ISIs are significant, indicating that it is possible the influence of covariates on transition dynamics found in \cite{sarkar2018bayesian} largely comes from ISIs. 

Moreover, the mixture gamma model for the ISIs that we proposed here may be of independent interest outside the scope of the Foxp2 application. 
The model is non-trivial and brought in additional statistical challenges, including posterior computation for unknown gamma  parameters and selection of the unknown number of mixture components. 
To our knowledge, sophisticated mixed effects mixture models for continuous variables that also simultaneously allow covariate selection have not been proposed in the literature before.

Our model is quite genetic in nature 
and hence can be used to analyze other data sets comprising categorical sequences and associated continuous inter-state interval times,  
both of which may potentially be influenced by various exogenous factors as well as subject-specific heterogeneity. 
{ 
The field of vocal communication neuroscience constitutes an important area of neuroscience research. 
Investigating scientific questions using animal models in controlled laboratory environments is a standard practice in this field. 
These researchers often use the standard two-way ANOVA design like the Foxp2 study analyzed in our manuscript. 
Our method is broadly applicable to such studies.  
We also cite below some examples from other application domains that can benefit from such analyses. 
In a study of asthma patients, \cite{combescure2003assessment} estimated the control states (optimal, sub-optimal or unacceptable) of 371 asthma patients with different BMI and disease severity over a four-year period. 
In an education assessment study, \cite{zhang2019scenario}  recorded sequences of writing states, characterized by keystroke logs, for  257 eighth graders of various genders, races and socioeconomic statuses. 
These works used Cox regression models to incorporate potential covariate influences but ignored the heterogeneity among the individuals \citep{dabrowska1994cox,krol2015semimarkov,guo2019writing}. 
To our knowledge, there exists no other statistical approach that flexibly accommodates population and individual-level effects in the transition as well as the inter-state interval distributions 
while also selecting the significant covariates for both. 
}

{Finally,  in this paper, we focused on discrete exogenous factors 
due to the nature of our motivating Foxp2 application. 
A simple but practically effective way to incorporate continuous covariates into the model is by categorizing them using, for example, their quantiles. 
Future research could investigate more principled ways to incorporate continuous covariates in our model.} 
The proposed model also points to possible directions of research in Markov renewal models 
that consider different aspects of state transitions (transition probabilities, inter-state intervals, state durations, etc.) 
and help make novel discoveries in important practical applications.

\baselineskip=17pt
\section*{Supplementary Materials}
The supplementary materials discuss the choice of the prior hyper-parameters, details of the MCMC sampler, additional results for the Foxp2 data set, and a simulation study. 
An \texttt{R} package \texttt{BMRMM} implementing our method is available at the Comprehensive R Archive Network (CRAN) and can also be accessed at \url{https://github.com/abhrastat/BMRMM}. 
A `readme' file is included here that instructs how to use the package to run the specific analysis reported here.

\baselineskip=14pt
\bibliographystyle{natbib}
\bibliography{NHMM,Neuroscience}

\begin{thebibliography}{}

\bibitem[Alvarez(2005)Alvarez]{alvarez2005estimation}
Alvarez, E.~E. (2005).
\newblock Estimation in stationary {M}arkov renewal processes, with application
  to earthquake forecasting in {T}urkey.
\newblock {\em Methodology and Computing in Applied Probability\/}, {\bf 7},
  119--130.

\bibitem[Arriaga and Jarvis(2013)Arriaga and Jarvis]{arriaga2013mouse}
Arriaga, G. and Jarvis, E.~D. (2013).
\newblock Mouse vocal communication system: Are ultrasounds learned or innate?
\newblock {\em Brain and Language\/}, {\bf 124}, 96--116.

\bibitem[Bulla and Muliere(2007)Bulla and Muliere]{bulla2007bayesian}
Bulla, P. and Muliere, P. (2007).
\newblock Bayesian nonparametric estimation for reinforced {M}arkov renewal
  processes.
\newblock {\em Statistical Inference for Stochastic Processes\/}, {\bf 10},
  283--303.

\bibitem[Castellucci {\em et~al.}(2016)Castellucci, McGinley, and
  McCormick]{Castellucci_etal:2016}
Castellucci, G.~A., McGinley, M.~J., and McCormick, D.~A. (2016).
\newblock Knockout of {F}oxp2 disrupts vocal development in mice.
\newblock {\em Nature Scientific Reports\/}, {\bf 6}.
\newblock doi: 10.1038/srep23305.

\bibitem[Chabout {\em et~al.}(2012)Chabout, Serreau, Ey, Bellier, Aubin,
  Bourgeron, and Granon]{Chabout_etal:2012}
Chabout, J., Serreau, P., Ey, E., Bellier, L., Aubin, T., Bourgeron, T., and
  Granon, S. (2012).
\newblock Adult male mice emit context-specific ultrasonic vocalizations that
  are modulated by prior isolation or group rearing environment.
\newblock {\em PLoS ONE\/}, {\bf 7:e29401}.
\newblock doi: 10.1371/journal.pone.0046610.

\bibitem[Chabout {\em et~al.}(2015)Chabout, Sarkar, Dunson, and
  Jarvis]{Chabout_etal:2015}
Chabout, J., Sarkar, A., Dunson, D.~B., and Jarvis, E.~D. (2015).
\newblock Male song syntax depends on contexts and influences female
  preferences in mice.
\newblock {\em Frontiers in Behavioral Neuroscience\/}, {\bf 9}, 1--19.

\bibitem[Chabout {\em et~al.}(2016)Chabout, Sarkar, Patel, Raiden, Dunson,
  Fisher, and Jarvis]{Chabout_etal:2016}
Chabout, J., Sarkar, A., Patel, S., Raiden, T., Dunson, D.~B., Fisher, S.~E.,
  and Jarvis, E.~D. (2016).
\newblock A {F}oxp2 mutation implicated in human speech deficits alters
  sequencing of ultrasonic vocalizations in adult male mice.
\newblock {\em Frontiers in Behavioral Neuroscience\/}, {\bf 10}, 1--18.

\bibitem[Chen(2000)Chen]{chen2000probability}
Chen, S.~X. (2000).
\newblock Probability density function estimation using gamma kernels.
\newblock {\em Annals of the Institute of Statistical Mathematics\/}, {\bf 52},
  471--480.

\bibitem[Chung and Dunson(2009)Chung and Dunson]{chung_dunson:2009}
Chung, Y. and Dunson, D.~B. (2009).
\newblock Nonparametric {B}ayes conditional distribution modeling with variable
  selection.
\newblock {\em \JASA\/}, {\bf 104}, 1646--1660.

\bibitem[Combescure {\em et~al.}(2003)Combescure, Chanez, Saint-Pierre, Daures,
  Proudhon, Godard, {\em et~al.}]{combescure2003assessment}
Combescure, C., Chanez, P., Saint-Pierre, P., Daures, J.~P., Proudhon, H.,
  Godard, P., {\em et~al.} (2003).
\newblock Assessment of variations in control of asthma over time.
\newblock {\em European Respiratory Journal\/}, {\bf 22}, 298--304.

\bibitem[Dabrowska {\em et~al.}(1994)Dabrowska, Sun, and
  Horowitz]{dabrowska1994cox}
Dabrowska, D.~M., Sun, G.-W., and Horowitz, M.~M. (1994).
\newblock Cox regression in a markov renewal model: an application to the
  analysis of bone marrow transplant data.
\newblock {\em Journal of the American Statistical Association\/}, {\bf 89},
  867--877.

\bibitem[Damsleth(1975)Damsleth]{damsleth1975conjugate}
Damsleth, E. (1975).
\newblock Conjugate classes for {G}amma distributions.
\newblock {\em Scandinavian Journal of Statistics\/}, {\bf 2}, 80--84.

\bibitem[Epifani {\em et~al.}(2014)Epifani, Ladelli, and
  Pievatolo]{epifani2014bayesian}
Epifani, I., Ladelli, L., and Pievatolo, A. (2014).
\newblock Bayesian estimation for a parametric {M}arkov renewal model applied
  to seismic data.
\newblock {\em Electronic Journal of Statistics\/}, {\bf 8}, 2264--2295.

\bibitem[Fisher and Scharff(2009)Fisher and Scharff]{fisher2009foxp2}
Fisher, S.~E. and Scharff, C. (2009).
\newblock Foxp2 as a molecular window into speech and language.
\newblock {\em Trends in Genetics\/}, {\bf 25}, 166--177.

\bibitem[Foucher {\em et~al.}(2005)Foucher, Mathieu, Saint-Pierre, Durand, and
  Daur{\`e}s]{foucher2005semi}
Foucher, Y., Mathieu, E., Saint-Pierre, P., Durand, J.-F., and Daur{\`e}s,
  J.-P. (2005).
\newblock A semi-{M}arkov model based on generalized {W}eibull distribution
  with an illustration for {HIV} disease.
\newblock {\em Biometrical Journal: Journal of Mathematical Methods in
  Biosciences\/}, {\bf 47}, 825--833.

\bibitem[Fujita {\em et~al.}(2008)Fujita, Tanabe, Shiota, Ueda, Suwa, Momoi,
  and Momoi]{Fujita_etal:2008}
Fujita, E., Tanabe, Y., Shiota, A., Ueda, M., Suwa, K., Momoi, M.~Y., and
  Momoi, T. (2008).
\newblock Ultrasonic vocalization impairment of {F}oxp2 ({R552H}) knockin mice
  related to speech-language disorder and abnormality of {P}urkinje cells.
\newblock {\em Proceedings of the National Academy of Sciences\/}, {\bf 105},
  3117--3122.

\bibitem[Gaub {\em et~al.}(2016)Gaub, Fisher, and Ehret]{Gaub_etal:2016}
Gaub, S., Fisher, S.~E., and Ehret, G. (2016).
\newblock Ultrasonic vocalizations of adult male {F}oxp2 mutant mice:
  behavioral contexts of arousal and emotion.
\newblock {\em Genes, Brain and Behavior\/}, {\bf 15}, 243--259.

\bibitem[Guo {\em et~al.}(2019)Guo, Zhang, Deane, and Bennett]{guo2019writing}
Guo, H., Zhang, M., Deane, P., and Bennett, R.~E. (2019).
\newblock Writing process differences in subgroups reflected in keystroke logs.
\newblock {\em Journal of Educational and Behavioral Statistics\/}, {\bf 44},
  571--596.

\bibitem[Hanson(2006)Hanson]{hanson2006modeling}
Hanson, T.~E. (2006).
\newblock Modeling censored lifetime data using a mixture of gammas baseline.
\newblock {\em Bayesian Analysis\/}, {\bf 1}, 575--594.

\bibitem[Holy and Guo(2005)Holy and Guo]{Holy_Guo:2005}
Holy, T.~E. and Guo, Z. (2005).
\newblock Ultrasonic songs of male mice.
\newblock {\em PLoS Biology\/}, {\bf 3}, 2177--2186.

\bibitem[Jarvis(2019)Jarvis]{jarvis2019evolution}
Jarvis, E.~D. (2019).
\newblock Evolution of vocal learning and spoken language.
\newblock {\em Science\/}, {\bf 366}, 50--54.

\bibitem[Kr{\'o}l and Saint-Pierre(2015)Kr{\'o}l and
  Saint-Pierre]{krol2015semimarkov}
Kr{\'o}l, A. and Saint-Pierre, P. (2015).
\newblock Semi{M}arkov: An {R} package for parametric estimation in multi-state
  semi-{M}arkov models.
\newblock {\em Journal of Statistical Software\/}, {\bf 66}, 1--16.

\bibitem[Lai {\em et~al.}(2001)Lai, Fisher, Hurst, Vargha-Khadem, and
  Monaco]{lai2001forkhead}
Lai, C.~S., Fisher, S.~E., Hurst, J.~A., Vargha-Khadem, F., and Monaco, A.~P.
  (2001).
\newblock A forkhead-domain gene is mutated in a severe speech and language
  disorder.
\newblock {\em Nature\/}, {\bf 413}, 519--523.

\bibitem[Levy(1954)Levy]{levy1954processus}
Levy, P. (1954).
\newblock Processus semi-{M}arkoviens.
\newblock In {\em Proceedings of International Congress of Mathematics\/}.

\bibitem[MacEachern(1999)MacEachern]{maceachern1999dependent}
MacEachern, S.~N. (1999).
\newblock Dependent nonparametric processes.
\newblock In {\em ASA proceedings of the section on Bayesian statistical
  science\/}, pages 50--55. Alexandria, Virginia: American Statistical
  Association.

\bibitem[Miller(2019)Miller]{miller2019fast}
Miller, J.~W. (2019).
\newblock Fast and accurate approximation of the full conditional for {G}amma
  shape parameters.
\newblock {\em Journal of Computational and Graphical Statistics\/}, {\bf 28},
  476--480.

\bibitem[Miller(1980)Miller]{miller1980bayesian}
Miller, R.~B. (1980).
\newblock Bayesian analysis of the two-parameter {G}amma distribution.
\newblock {\em Technometrics\/}, {\bf 22}, 65--69.

\bibitem[Moles {\em et~al.}(2007)Moles, Costantini, Garbugino, Zanettini, and
  D\'Amato]{Moles_etal:2007}
Moles, A., Costantini, F., Garbugino, L., Zanettini, C., and D\'Amato, F.~R.
  (2007).
\newblock Ultrasonic vocalizations emitted during dyadic interactions in female
  mice: A possible index of sociability?
\newblock {\em Behavioural Brain Research\/}, {\bf 182}, 223–--230.

\bibitem[Mooney(2020)Mooney]{mooney2020neurobiology}
Mooney, R. (2020).
\newblock The neurobiology of innate and learned vocalizations in rodents and
  songbirds.
\newblock {\em Current opinion in neurobiology\/}, {\bf 64}, 24--31.

\bibitem[Muliere {\em et~al.}(2003)Muliere, Secchi, and
  Walker]{muliere2003reinforced}
Muliere, P., Secchi, P., and Walker, S.~G. (2003).
\newblock Reinforced random processes in continuous time.
\newblock {\em Stochastic Processes and their Applications\/}, {\bf 104},
  117--130.

\bibitem[M{\"u}ller {\em et~al.}(2011)M{\"u}ller, Quintana, and
  Rosner]{muller2011product}
M{\"u}ller, P., Quintana, F., and Rosner, G.~L. (2011).
\newblock A product partition model with regression on covariates.
\newblock {\em Journal of Computational and Graphical Statistics\/}, {\bf 20},
  260--278.

\bibitem[Musolf {\em et~al.}(2015)Musolf, Meindl, Larsen, Kalcounis-Rueppell,
  and J.]{Musolf_etal:2015}
Musolf, K., Meindl, S., Larsen, A.~L., Kalcounis-Rueppell, M.~C., and J., P.~D.
  (2015).
\newblock Ultrasonic vocalizations of male mice differ among species and
  females show assortative preferences for male calls.
\newblock {\em PLoS ONE\/}, {\bf 10:e0134123}.
\newblock doi:10.1371/journal.pone.0134123.

\bibitem[{NIH-NIDCD Report}(2020){NIH-NIDCD Report}]{NIH-NIDCD:2020}
{NIH-NIDCD Report} (2020).
\newblock Statistics on voice, speech, and language.
\newblock
  \url{https://www.nidcd.nih.gov/health/statistics/statistics-voice-speech-and-language}.

\bibitem[Phelan(1990)Phelan]{phelan1990bayes}
Phelan, M.~J. (1990).
\newblock Bayes estimation from a {M}arkov renewal process.
\newblock {\em The Annals of Statistics\/}, {\bf 18}, 603--616.

\bibitem[Pyke(1961)Pyke]{pyke1961markov}
Pyke, R. (1961).
\newblock Markov renewal processes: definitions and preliminary properties.
\newblock {\em The Annals of Mathematical Statistics\/}, {\bf 32}, 1231--1242.

\bibitem[Sarkar {\em et~al.}(2018)Sarkar, Chabout, Macopson, Jarvis, and
  Dunson]{sarkar2018bayesian}
Sarkar, A., Chabout, J., Macopson, J.~J., Jarvis, E.~D., and Dunson, D.~B.
  (2018).
\newblock Bayesian semiparametric mixed effects {M}arkov models with
  application to vocalization syntax.
\newblock {\em Journal of the American Statistical Association\/}, {\bf 113},
  1515--1527.

\bibitem[Scattoni {\em et~al.}(2011)Scattoni, Ricceri, and
  Crawley]{Scattoni_etal:2011}
Scattoni, M.~L., Ricceri, L., and Crawley, J.~N. (2011).
\newblock Unusual repertoire of vocalizations in adult {BTBR} {T}+tf/{J} mice
  during three types of social encounters.
\newblock {\em Genes, Brain and Behavior\/}, {\bf 10}, 44--56.

\bibitem[Smith(1955)Smith]{smith1955regenerative}
Smith, W.~L. (1955).
\newblock Regenerative stochastic processes.
\newblock {\em Proceedings of the Royal Society of London. Series A.
  Mathematical and Physical Sciences\/}, {\bf 232}, 6--31.

\bibitem[Vargha-Khadem {\em et~al.}(1998)Vargha-Khadem, Watkins, Price,
  Ashburner, Alcock, Connelly, Frackowiak, Friston, Pembrey, Mishkin, {\em
  et~al.}]{vargha1998neural}
Vargha-Khadem, F., Watkins, K.~E., Price, C., Ashburner, J., Alcock, K.~J.,
  Connelly, A., Frackowiak, R.~S., Friston, K.~J., Pembrey, M., Mishkin, M.,
  {\em et~al.} (1998).
\newblock Neural basis of an inherited speech and language disorder.
\newblock {\em Proceedings of the National Academy of Sciences\/}, {\bf 95},
  12695--12700.

\bibitem[Weiss and Zelen(1965)Weiss and Zelen]{weiss1965semi}
Weiss, G.~H. and Zelen, M. (1965).
\newblock A semi-{M}arkov model for clinical trials.
\newblock {\em Journal of Applied Probability\/}, {\bf 2}, 269--285.

\bibitem[Wiper {\em et~al.}(2001)Wiper, Insua, and Ruggeri]{wiper2001mixtures}
Wiper, M., Insua, D.~R., and Ruggeri, F. (2001).
\newblock Mixtures of gamma distributions with applications.
\newblock {\em Journal of computational and graphical statistics\/}, {\bf 10},
  440--454.

\bibitem[Wu and Ghosal(2008)Wu and Ghosal]{wu2008kullback}
Wu, Y. and Ghosal, S. (2008).
\newblock Kullback-{L}eibler property of kernel mixture priors in {B}ayesian
  density estimation.
\newblock {\em Electronic Journal of Statistics\/}, {\bf 2}, 298--331.

\bibitem[Yang and Hursch(1973)Yang and Hursch]{yang1973use}
Yang, M.~C. and Hursch, C.~J. (1973).
\newblock The use of a semi-{M}arkov model for describing sleep patterns.
\newblock {\em Biometrics\/}, {\bf 29}, 667--676.

\bibitem[Zhang {\em et~al.}(2019)Zhang, van Rijn, Deane, and
  Bennett]{zhang2019scenario}
Zhang, M., van Rijn, P.~W., Deane, P., and Bennett, R.~E. (2019).
\newblock Scenario-based assessments in writing: An experimental study.
\newblock {\em Educational Assessment\/}, {\bf 24}, 73--90.

\end{thebibliography}

\clearpage\pagebreak\newpage
\pagestyle{fancy}
\fancyhf{}
\rhead{\bfseries\thepage}
\lhead{\bfseries SUPPLEMENTARY MATERIALS}

\baselineskip 20pt
\vspace{-0.5cm}
\begin{center}
{\LARGE{Supplementary Materials for}\\ 
{\bf Bayesian Semiparametric \\ Markov Renewal Mixed Models
for Vocalization Syntax \\
}}
\end{center}

\setcounter{equation}{0}
\setcounter{page}{1}
\setcounter{table}{1}
\setcounter{figure}{0}
\setcounter{section}{0}
\numberwithin{table}{section}
\renewcommand{\theequation}{S.\arabic{equation}}
\renewcommand{\thesubsection}{S.\arabic{section}.\arabic{subsection}}
\renewcommand{\thesection}{S.\arabic{section}}
\renewcommand{\thepage}{S.\arabic{page}}
\renewcommand{\thetable}{S.\arabic{table}}
\renewcommand{\thefigure}{S.\arabic{figure}}
\baselineskip=12pt

\begin{center}
Yutong Wu\\
Department of Mechanical Engineering\\
The University of Texas at Austin\\
204 E. Dean Keeton Street C2200, Austin, TX 78712-1591, USA\\
yutong.wu@utexas.edu\\
\vskip 10pt 
Erich D. Jarvis\\
The Rockefeller University, New York, NY 10065, USA\\
Howard Hughes Medical Institute, Chevy Chase, MD 20815, USA\\
ejarvis@rockefeller.edu \\
\vskip 10pt 
Abhra Sarkar\\
Department of Statistics and Data Sciences\\ 
The University of Texas at Austin\\ 
2317 Speedway D9800, Austin, TX 78712-1823, USA\\
abhra.sarkar@utexas.edu\\ 
\vskip 10pt 
\end{center}

\baselineskip=15pt

\vskip 10pt 
{The supplementary materials provide additional information on transition types in Appendix~\ref{sec: data set info}, details of the choice of the hyper-parameters in Appendix~\ref{sec: hyper-parameters}, MCMC initialization in Appendix~\ref{sec: initialization}, and a step-by-step posterior computation in Appendix~\ref{sec: pc}.
We discuss the need for mouse-specific mixture coefficients in Appendix~\ref{sec:coeff}. 
{Finally, we summarize the results of a simulation study in Appendix~\ref{sec: sim study}}.
The \texttt{R} code and the corresponding instructions are provided in Supplemental Materials for interested readers.}

\newpage

\section{Additional  Information  on the Data Set} \label{sec: data set info}

{Recall that the Foxp2 data set contains 70818 rows, including 49 songs sung by 18 mice with different genotypes ($F$ or $W$) under one of the three social contexts ($U$, $L$ or $A$). 
The song consists of syllables, which can be of type \texttt{d},  \texttt{m}, \texttt{s},  or \texttt{u}.
We present the transition types stratified by genotypes in Table~\ref{tab: transition types genotype} and by social contexts in Table~\ref{tab: transition types context}. }
\begin{table}[ht]
\parbox{.45\linewidth}{
\centering
\caption*{Genotype $F$}
\vspace{-2ex}
\begin{tabular}{c*{4}r}
 \multicolumn{1}{c}{} & \myalign{c}{\texttt{d}} & \myalign{c}{\texttt{m}} & \myalign{c}{\texttt{s}} & \myalign{c}{\texttt{u}}  \\ \hline
\texttt{d} 	& 1101         & 336       & 2444           & 94        	 \\
\texttt{m} 	& 366          & 399        & 1282           & 91          	\\
\texttt{s} 	& 2381         & 1317       & 20629          & 757       \\
\texttt{u} 	& 125          & 89        & 729            & 42          		\\
\hline
\end{tabular}
}
\quad~~~
\parbox{.45\linewidth}{
\centering
\caption*{Genotype $W$}
\vspace{-2ex}
\begin{tabular}{c*{4}r}
 \multicolumn{1}{c}{} & \myalign{c}{\texttt{d}} & \myalign{c}{\texttt{m}} & \myalign{c}{\texttt{s}} & \myalign{c}{\texttt{u}}  \\ \hline
\texttt{d} 	& 1679          & 628       & 3536           & 174        	 \\
\texttt{m} 	& 621          & 584        & 1905           & 166          	\\
\texttt{s} 	& 3539         & 1896       & 21509          & 985       \\
\texttt{u} 	& 180          & 168        & 976            & 90          		\\
\hline
\end{tabular}
}
\caption{{Empirical distributions of the syllable transitions stratified by genotype. }}
\label{tab: transition types genotype}
\end{table}
\begin{table}[ht]
\hspace{4cm}
\parbox{.3\linewidth}{
\centering
\caption*{Context $U$}
\vspace{-2ex}
\begin{tabular}{c*{4}r}
 \multicolumn{1}{c}{} & \myalign{c}{\texttt{d}} & \myalign{c}{\texttt{m}} & \myalign{c}{\texttt{s}} & \myalign{c}{\texttt{u}}  \\ \hline
\texttt{d} 	& 613         & 153       & 977           & 31        	 \\
\texttt{m} 	& 158          & 142        & 350           & 22          	\\
\texttt{s} 	& 961         & 348       & 6349          & 180       \\
\texttt{u} 	& 44          & 28        & 161            & 8          		\\
\hline
\end{tabular}
}\\[6pt]
\parbox{.45\linewidth}{
\centering
\caption*{Context $L$}
\vspace{-2ex}
\begin{tabular}{c*{4}r}
 \multicolumn{1}{c}{} & \myalign{c}{\texttt{d}} & \myalign{c}{\texttt{m}} & \myalign{c}{\texttt{s}} & \myalign{c}{\texttt{u}}  \\ \hline
\texttt{d} 	& 1001         & 549       & 3019           & 181        	 \\
\texttt{m} 	& 577          & 684        & 2171           & 205          	\\
\texttt{s} 	& 2979         & 2214       & 24627          & 1228       \\
\texttt{u} 	& 190          & 194        & 1229            & 104          		\\
\hline
\end{tabular}
}
\parbox{.45\linewidth}{
\centering
\caption*{Context $A$}
\vspace{-2ex}
\begin{tabular}{c*{4}r}
 \multicolumn{1}{c}{} & \myalign{c}{\texttt{d}} & \myalign{c}{\texttt{m}} & \myalign{c}{\texttt{s}} & \myalign{c}{\texttt{u}}  \\ \hline
\texttt{d} 	& 1166          & 262       & 1984           & 56        	 \\
\texttt{m} 	& 252          & 157        & 666           & 30          	\\
\texttt{s} 	& 1980         & 651       & 11162          & 334       \\
\texttt{u} 	& 71          & 35        & 315            & 20          		\\
\hline
\end{tabular}
}
\caption{{Empirical distributions of the syllable transitions stratified by context. }}
\label{tab: transition types context}
\end{table}
\section{Choice of Hyper-Parameters}\label{sec: hyper-parameters}

\subsection{Hyper-Parameters for Syllable Transitions}

For the inference of the transition of syllables, we set $\alpha_{trans,00}=1$ and $\lambda_{trans,00}(y_{t})=\sum_{s,t}1\{y_{s,t}=y_{t}\}/\sum_{s}T_{s}$, the overall percentage of syllables among all songs. 
We set each $\alpha_{trans,j}$ at the value for which $p_{0}(H_{0,trans,j})=p_{0}(k_{trans,j}=1)=1/2$.  
For the remaining fixed hyper-parameters, we set $a_{trans,0}=a_{trans,1}=b_{trans,0}=a_{trans}^{(0)}=b_{trans}^{(0)}=1$. 
In numerical experiments, results were robust to these choices.

\subsection{Hyper-Parameters for Inter-Syllable Intervals}

For the inference of the inter-syllable intervals, $\alpha_{isi,00}$ is set to $1$ and the global probability vector $\blambda_{isi,00}(\cdot)$ is the ratio of the component size, determined via $k$-means, over the total sample size. 
Similar to the syllable transitions, we set $a_{isi,0}=a_{isi,1}=b_{isi,0}=a_{isi}^{(0)}=b_{isi}^{(0)}=1$, 
and, as before, the results were seen to be robust to these choices.

Now we discuss how we determine the number of mixture components, $K$.
To find a suitable choice for $K$, one use the log pseudo marginal likelihood (LPML) (Geisser and Eddy, 1979)  
or the widely applicable information criterion (WAIC) (Watanabe and Opper, 2010), 
which are easily calculated and hence particularly useful for complex Bayesian hierarchical models like ours.
Specifically, let $f(\wt\tau_{s,t} \mid \alpha,\beta,\pi)$ be the mixture gamma density function for the transformed ISI $\wt\tau_{s,t}$ with shape and rate parameters $\alpha,\beta$ and mixture probabilities $\pi$. 
The LPML and the WAIC are then defined as follows.
\bse
&& LPML = -\sum_{s,t} \log \left[\mathbb{E}_{\alpha,\beta,\pi \mid \wt\btau} \left\{\frac{1}{f(\wt\tau_{s,t} \mid \alpha,\beta,\pi)}\right\}\right], \\
&& WAIC = -2\left[\sum_{s,t} \log \> \mathbb{E}_{\alpha,\beta,\pi \mid \wt\btau}\left\{f(\wt\tau_{s,t} \mid \alpha,\beta,\pi) \right\} - p_{WAIC}\right], \\
&& \text{ where } \> p_{WAIC} = 2\sum_{s,t} \left[\log\> \mathbb{E}_{\alpha,\beta,\pi \mid \wt\btau}\left\{f(\wt\tau_{s,t} \mid \alpha,\beta,\pi) \right\} -  \mathbb{E}_{\alpha,\beta,\pi \mid \wt\btau}\left\{\log \> f(\wt\tau_{s,t} \mid \alpha,\beta,\pi) \right\} \right].
\ese
They can be straightforwardly estimated using the MCMC samples of $\alpha, \beta, \pi$. 

\begin{figure}[ht]
\centering
\subfloat[LPML scores for different $K$'s.]{
\includegraphics[width=0.45\textwidth]{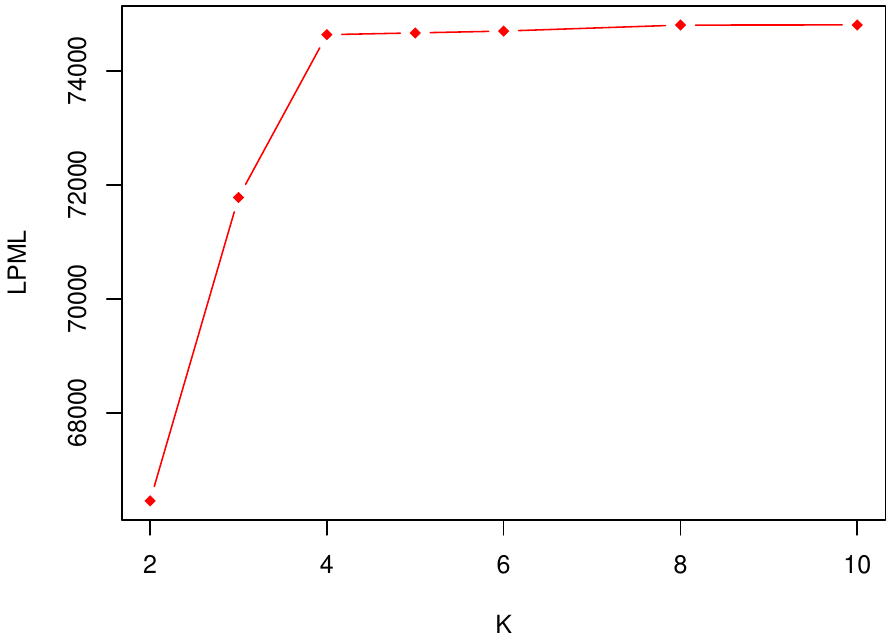}
\label{fig:lpml}}
\qquad
\subfloat[WAIC scores for different $K$'s.]{
\includegraphics[width=0.45\textwidth]{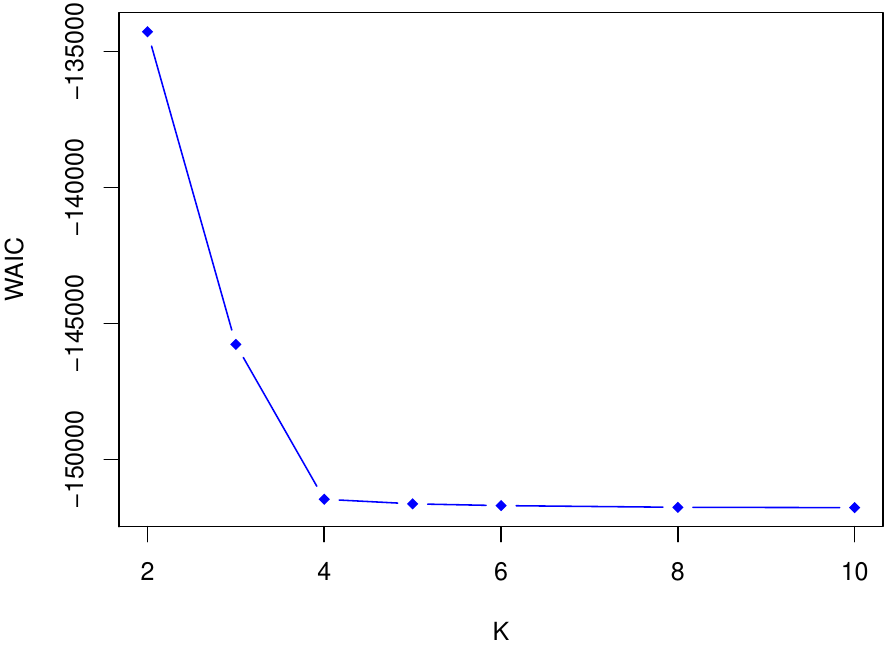}
\label{fig:waic}}
\caption{Results of the model comparison for $K=2,3,4,5,6,8,10$, including LPML (left) and WAIC (right) scores. 
Larger values of LPML and smaller values of WAIC indicate better model fits.}
\label{fig:model_selection}
\end{figure}

Figure~\ref{fig:model_selection} displays the LPML (left) and WAIC (right) for mixture gamma distributions with $K \in \{2,3,4,5,6,8,10\}$ mixture components. 
Larger values of LPML and smaller values of WAIC indicate better model fits.
We see that $K=2$ and $K=3$ have much worse scores compared to $K\ge 4$, suggesting that three or fewer components are highly insufficient. 
Starting at $K=4$, both the LPML and WAIC scores converge, suggesting that having more than four components does not improve the fit significantly. 
By Occam's razor principle, among the models with similar fit, we choose the simplest one, that is, $K=4$.

\section{MCMC Initialization} \label{sec: initialization}


\subsection{Initialization for Syllable Transitions}\label{sec: prior hyper trans}

For the two covariates, genotype and social contexts, we initialize $\bmu_{trans,1}=(1/2, 1/2)\trans$ and $\bmu_{trans,2}=(1/3, 1/3, 1/3)\trans$. 
We assume, at the beginning of the simulation, that each covariate value has its own cluster and initialize $z_{trans,j,h}$ at $h$ for $h=1,\dots,d_{j}$. 
Accordingly, the associated $\blambda_{trans,h_{1},h_{2}}(y_{t} \mid y_{t-1})$ are initialized at $\sum_{s,t}1\{y_{s,t}=y_{t},y_{s,t-1}=y_{t-1},x_{s,1}={h}_{1},x_{s,2}={h}_{2}\}/\sum_{s,t}1\{y_{s,t-1}=y_{t-1},x_{s,1}={h}_{1},x_{s,2}={h}_{2}\}$. 
The individual effect $\blambda^{(i)}(y_{t} \mid y_{t-1})$ is initialized at $\sum_{s,t}1\{y_{s,t}=y_{t},y_{s,t-1}=y_{t-1},i_{s}=i\}/\sum_{s,t}1\{y_{s,t-1}=y_{t-1},i_{s}=i\}$. 
For each $y_{t-1}$ and $i$, $\{\pi^{(i)}_{trans,0}(y_{t-1}),\pi^{(i)}_{trans,1}(y_{t-1})\}$ is initialized at $(0.8,0.2)$, indicating a higher initial probability for population-level effect.

\subsection{Initialization for Inter-Syllable Intervals}\label{sec: prior hyper isi}

For the gamma mixtures, we use $k$-means to form the initial $K=4$ components and set $z_{isi,s,t}$ as the label of the component based on the results of $k$-means.
Similar to syllable transitions, we initialize $\{\pi^{(i)}_{isi,0}(k),\pi^{(i)}_{isi,1}(k)\}$ as $(0.8,0.2)$ for each $i$ and $k=1,\dots,K$. 
We assume, initially, that each level of $x_r$ for covariate $r$ forms its own clusters and initialize the latent cluster allocation variables $z_{isi,r,w}$ at $w$ for $w=1,\dots,d_{r}$ for $r=1,2,3$.
{Then the population-level mixture probability vector is set to $\blambda_{isi,g_1,g_2,g_3}(k)=\sum_{s,t} 1\{z_{isi,s,t}=k,x_{s,1}={g}_{1},x_{s,2}={g}_{2}, {(y_{s,t-1},y_{s,t})}=g_3\}/\sum_{s,t}1\{x_{s,1}={g}_{1},x_{s,2}={g}_{2}, {(y_{s,t-1},y_{s,t})}=g_3\}$ for $k=1,\dots,K$.} 
The individual-level mixture probability vector $\blambda_{isi}^{(i)}(k)$ is initialized at $\sum_{s,t} 1\{z_{isi,s,t}=k,i_{s}=i\}/\sum_{s,t}1\{i_s=i\}$ for $k=1,\dots,K$.
The mean probability vector $\blambda_{isi,0}$ is initialized at $\blambda_{isi,00}$.

\section{Posterior Computation}\label{sec: pc}
Samples are drawn from the posterior using a Gibbs sampler that exploits the conditional independence relationships 
depicted in (2) and (5) in the main paper.
For convenience, we use a generic variable $\bzeta$ to denote all other variables, including data points. 
In the following two sub-sections, we detail the step-by-step posterior computation for each variable of the syllable transitions and ISIs, respectively.

\subsection{Posterior Computation for Syllable Transitions}\label{sec: pc trans}

\begin{enumerate}
\item 
Sample each $z_{trans,j,\ell}$ according to its multinomial full conditional  
\bse
&& p(z_{trans,j,\ell}=h_{j} \mid z_{trans,j',\ell}=h_{j'}, j'\neq j, \bzeta) \propto \mu_{trans,h_{j}}^{(j)}  \times \\
&& \resizebox{\linewidth}{!}{$\prod_{y_{t-1}}\prod_{(h_{1},h_{2})}  \frac{\beta\{\alpha_{trans,0}\lambda_{trans,0}(1 \mid y_{t-1})+n_{h_{1},h_{2}}(1 \mid y_{t-1}), \dots, \alpha_{trans,0}\lambda_{trans,0}(4 \mid y_{t-1})+n_{h_{1},h_{2}}(4 \mid y_{t-1})\}}{\beta\{\alpha_{trans,0}\lambda_{trans,0}(1 \mid y_{t-1}),\dots,\alpha_{trans,0}\lambda_{trans,0}(4 \mid y_{t-1})\}}$,}
\ese
where $n_{h_{1},h_{2}}(y_{t} \mid y_{t-1}) = \sum_{s,t}1\{y_{s,t}=y_{t}, y_{s,t-1}=y_{t-1},v_{trans,s,t}=0, z_{trans,1,x_{s,1}}=h_{1},z_{trans,2,x_{s,2}}=h_{2}\}$. 
Note that $v_{trans,s,t}\in\{0,1\}$'s are latent variables for $y_{s,t}$'s, indicating if the transition probability follows the population-level effect ($v_{trans,s,t}=0$) or the individual-level effect ($v_{trans,s,t}=1$), as in \cite{sarkar2018bayesian}.

\item 
Sample each $\bmu_{trans,j}$ according to its Dirichlet full conditional  
\bse
&& \{\mu_{trans,j}(1),\dots,\mu_{trans,j}(d_{j})\} \mid \bzeta \sim \Dir\left\{\alpha_{trans,j}+n_{j}(1),\dots,\alpha_{trans,j}+n_{j}(d_{j})\right\},
\ese
where $n_{j} (h)= \sum_{\ell=1}^{d_{j}}1\{z_{trans,j,\ell}=h\}$. 

\item 
Sample each $v_{trans,s,t}$ according to its Bernoulli full conditional  
\bse
&& p(v_{trans,s,t}=v \mid i_s=i, \bzeta) \propto \pi_{trans,v}^{(i)}(y_{s,t-1})  \times \wt\lambda_{trans,v}(y_{s,t} \mid y_{s,t-1}),
\ese
where $\wt\blambda_{trans,0}(\cdot \mid y_{t-1})=\blambda_{trans,h_{1},h_{2}}(\cdot \mid y_{t-1})$ with $(z_{trans,1,x_{s,1}},z_{trans,2,x_{s,2}})=(h_{1},h_{2})$, and $\wt\blambda_{trans,1}(\cdot \mid y_{t-1})=\blambda^{(i)}_{trans}(\cdot \mid y_{t-1})$.

\item 
Sample $\bpi^{(i)}_{trans} = \{\pi_{trans,0}^{(i)}(y_{t-1}), \pi_{trans,1}^{(i)}(y_{t-1})\}\trans$'s according to its Beta full conditional  
\bse
&& \{\pi_{trans,0}^{(i)}(y_{t-1}),\pi_{trans,1}^{(i)}(y_{t-1})\} \mid \bzeta \sim \Beta\left\{a_{trans,0}+n_{0}^{(i)}(y_{t-1}),a_{trans,1}+n_{1}^{(i)}(y_{t-1})\right\},
\ese
where $n_{v}^{(i)} (y_{t-1})= \sum_{s,t}1\{v_{trans,s,t}=v, y_{s,t-1}=y_{t-1}, i_s=i\}$. 

\item 
Sample each $\blambda^{(i)}_{trans}(\cdot \mid y_{t-1})$'s according to its Dirichlet full conditional 
\bse
&& \hspace{-1cm} \{\lambda^{(i)}_{trans}(1 \mid y_{t-1}),\dots,\lambda^{(i)}_{trans}(4 \mid y_{t-1})\} \mid \bzeta \sim \\
&&	\Dir\left\{\alpha^{(0)}_{trans} \lambda_{trans,0}(1 \mid y_{t-1})+n^{(i)}(1 \mid y_{t-1}),\dots,\alpha_{trans}^{(0)} \lambda_{trans,0}(4 \mid y_{t-1})+n^{(i)}(4 \mid y_{t-1})\right\},
\ese
where $n^{(i)}(y_{t} \mid y_{t-1}) = \sum_{s,t}1\{y_{s,t}=y_{t}, y_{s,t-1}=y_{t-1}, v_{trans,s,t}=1, i_{s}=i\}$.

\item 
Sample each $\blambda_{trans,h_{1},h_{2}}(\cdot \mid y_{t-1})$ according to its Dirichlet full conditional 
\bse
&& \hspace{-1cm} \{\lambda_{trans,h_{1},h_{2}}(1 \mid y_{t-1}),\dots,\lambda_{trans,h_{1},h_{2}}(4 \mid y_{t-1})\} \mid \bzeta \sim \\
&&	\Dir\left\{\alpha_{trans,0}\lambda_{trans,0}(1 \mid y_{t-1})+n_{h_{1},h_{2}}(1 \mid y_{t-1}),\dots,\alpha_{trans,0}\lambda_{trans,0}(4 \mid y_{t-1})+n_{h_{1},h_{2}}(4 \mid y_{t-1})\right\},
\ese
where $n_{h_1,h_2}(y_{t} \mid y_{t-1}) = \sum_{s,t}1\{y_{s,t}=y_{t}, y_{s,t-1}=y_{t-1}, v_{trans,s,t}=0, z_{trans,1,x_{1}}=h_1, z_{trans,2,x_2}=h_2\}$.

\item 
For $\ell=n_{h_{1},h_{2}}(y_{t} \mid y_{t-1})$, 
sample auxiliary variables $v_{\ell,h_{1},h_{2}}(y_{t} \mid y_{t-1})$ 
\bse
v_{\ell,h_{1},h_{2}}(y_{t} \mid y_{t-1}) \mid \bzeta \sim \Bern\left\{\frac{\alpha_{trans,0}\lambda_{trans,0}(y_{t} \mid y_{t-1})}{\ell-1+\alpha_{trans,0}\lambda_{trans,0}(y_{t} \mid y_{t-1})}\right\}.
\ese
Set $v_{h_{1},h_{2}}(y_{t} \mid y_{t-1})=\sum_{\ell} v_{\ell,h_{1},h_{2}}(y_{t} \mid y_{t-1})$. 
Likewise, for $\ell=1,\dots,n^{(i)}(y_{t} \mid y_{t-1})$, sample auxiliary variables $v_{\ell}^{(i)}(y_{t} \mid y_{t-1})$ as
\bse
v_{\ell}^{(i)}(y_{t} \mid y_{t-1}) \mid \bzeta \sim \Bern\left\{\frac{\alpha_{trans}^{(0)}\lambda_{trans,0}(y_{t} \mid y_{t-1})}{\ell-1+\alpha^{(0)}_{trans}\lambda_{trans,0}(y_{t} \mid y_{t-1})}\right\}.
\ese
Set $v^{(i)}(y_{t} \mid y_{t-1})=\sum_{\ell}v_{\ell}^{(i)}(y_{t} \mid y_{t-1})$. 
Additionally, sample auxiliary variables 
\bse
r_{h_{1},h_{2}}(y_{t-1}) \mid \bzeta \sim \Beta\{\alpha_{trans,0}+1,n_{h_{1},h_{2}}(y_{t-1})\}, \\
s_{h_{1},h_{2}}(y_{t-1}) \mid \bzeta \sim \Bern\left\{\frac{n_{h_{1},h_{2}}(y_{t-1})}{n_{h_{1},h_{2}}(y_{t-1})+\alpha_{trans,0}}\right\}, \\
r^{(i)}(y_{t-1}) \mid \bzeta \sim \Beta\{\alpha^{(0)}_{trans}+1,n^{(i)}(y_{t-1})\}, \\
s^{(i)}(y_{t-1}) \mid \bzeta \sim \Bern\left\{\frac{n^{(i)}(y_{t-1})}{n^{(i)}(y_{t-1})+\alpha_{trans}^{(0)}}\right\},
\ese
where $n_{h_{1},h_{2}}(y_{t-1}) = \sum_{y_{t}} n_{h_{1},h_{2}}(y_{t} \mid y_{t-1}) $ and $n^{(i)}(y_{t-1})=\sum_{y_{t}}n^{(i)}(y_{t} \mid y_{t-1})$.
Set 
$v(y_{t} \mid y_{t-1}) = \sum_{h_{1},h_{2}}v_{h_{1},h_{2}}(y_{t} \mid y_{t-1}) + \sum_{i} v^{(i)}(y_{t} \mid y_{t-1})$.
Also, set $v_{0}=\sum_{y_{t}}\sum_{y_{t-1}}\sum_{h_{1},h_{2}} v_{h_{1},h_{2}}(y_{t} \mid y_{t-1})$,  
$v^{(0)}=\sum_{y_{t}}\sum_{y_{t-1}}\sum_{i}v^{(i)}(y_{t} \mid y_{t-1})$, 
$\log ~ r_{0}=\sum_{y_{t-1}}\sum_{h_{1},h_{2}} \log~r_{h_{1},h_{2}}(y_{t-1})$, 
$s_{0}=\sum_{y_{t-1}}\sum_{h_{1},h_{2}}s_{h_{1},h_{2}}(y_{t-1})$, 
$\log ~ r^{(0)}=\sum_{y_{t-1}}\sum_{i} \log~r^{(i)}(y_{t-1})$, and 
$s^{(0)}=\sum_{y_{t-1}}\sum_{i}s^{(i)}(y_{t-1})$.

\item 
Sample $\alpha_{trans,0}$ and $\alpha^{(0)}_{trans}$ according to their gamma full conditionals 
\bse
&& \alpha_{trans,0} \mid \bzeta \sim \Ga (a_{trans,0}+v_{0}-s_{0}, b_{trans,0}-\log~r_{0}), \\ 
&& \alpha^{(0)}_{trans} \mid \bzeta \sim \Ga (a^{(0)}_{trans}+v^{(0)}-s^{(0)}, b^{(0)}_{trans}- \log~r^{(0)}). 
\ese

\item 
~Finally, sample $\blambda_{trans,0}$ according to its Dirichlet full conditional 
\bse
&& \hspace{-1cm} \{\lambda_{trans,0}(1 \mid y_{t-1}),\dots,\lambda_{trans,0}(4 \mid y_{t-1})\} \mid \bzeta \sim \\
&& \Dir\{\alpha_{trans,00}\lambda_{trans,00}(1)+v(1 \mid y_{t-1}),\dots,\alpha_{trans,00}\lambda_{trans,00}(4)+v(4 \mid y_{t-1})\}. 
\ese
\end{enumerate}

The steps to update the hyper-parameters $\alpha_{trans,0}$, $\alpha^{(0)}_{trans}$ and the global transition distributions $\blambda_{trans,0}$ followed the auxiliary variable sampler in West (1992) and Teh~{\em et al.} (2006).

\subsection{Posterior Computation for Inter-Syllable Intervals}\label{sec:pc isi}

\begin{enumerate}

\item Sample the mixture component latent variable $z_{isi,s,t}$'s 
\bse
&& {p(z_{isi,s,t}=k \mid i_s= i, z_{isi,1,x_{1}}=g_{1},z_{isi,2,x_{2}}=g_{2},z_{isi,3,{(y_{s,t-1},y_{s,t})}}=g_{3}, \bzeta)}\nonumber \\ 
&& ~~~~~~\propto P_{isi,g_{1},g_{2},g_{3}}^{(i)}(k) \times Ga(\tau_{s,t} \mid \alpha_{k}, \beta_{k}),
\ese

where $P_{isi,g_{1},g_{2},g_{3}}^{(i)}(k)=\pi_{isi,0}^{(i)}\blambda_{isi,g_{1},g_{2},g_{3}}(k)+\pi_{isi,1}^{(i)}\blambda_{isi}^{(i)}(k)$ for $k=1,\dots,K$.

\item 
Similar to transition probabilities, we introduce $v_{isi,s,t}$'s and sample each $v_{isi,s,t}$ according to its Bernoulli full conditional  
\bse
&& p(v_{isi,s,t}=v \mid i_s=i, \bzeta) \propto \pi_{isi,v}^{(i)}(k)  \times \wt\lambda_{isi,v}(k),
\ese
where $\wt\blambda_{isi,0}(\cdot)=\blambda_{isi,g_{1},g_{2},g_{3}}(\cdot)$ with {$(z_{isi,1,x_{1}},z_{isi,2,x_{2}},z_{isi,3,{(y_{s,t-1},y_{s,t})}})=(g_{1},g_{2},g_{3})$}, $\wt\blambda_{isi,1}(\cdot)=\blambda^{(i)}_{isi}(\cdot)$.

\item 
Sample $\bpi^{(i)}_{isi} = \{\pi_{isi,0}^{(i)}(k), \pi_{isi,1}^{(i)}(k)\}\trans$'s according to its Beta full conditional  
\bse
&& \{\pi_{isi,0}^{(i)}(k),\pi_{isi,1}^{(i)}(k)\} \mid \bzeta \sim \Beta\left\{a_{isi,0}+n_{0}^{(i)}(k),a_{isi,1}+n_{1}^{(i)}(k)\right\},
\ese
where $n_{v}^{(i)} (k)= \sum_{s,t}1\{z_{isi,s,t}=k, v_{isi,s,t}=v, i_s=i\}$.

\item 
Sample each $\blambda^{(i)}_{isi}(\cdot)$'s according to its Dirichlet full conditional 
\bse
&& \hspace{-1cm} \{\lambda^{(i)}_{isi}(1),\dots,\lambda^{(i)}_{isi}(K)\} \mid \bzeta \sim \Dir\left\{\alpha^{(0)}_{isi} \lambda_{isi,0}(1)+n^{(i)}(1),\dots,\alpha^{(0)}_{isi} \lambda_{isi,0}(K)+n^{(i)}(K)\right\},
\ese
where $n^{(i)}(k) = \sum_{s,t}1\{z_{isi,s,t}=k, v_{isi,s,t}=1, i_{s}=i\}$.

\item 
Sample each $\blambda_{isi,g_{1},g_{2},g_{3}}(\cdot)$ according to its Dirichlet full conditional 
\bse
&& \hspace{-1cm} \{\lambda_{isi,g_{1},g_{2},g_{3}}(1),\dots,\lambda_{isi,g_{1},g_{2},g_{3}}(K)\} \mid \bzeta \sim \\
&&	\Dir\{\alpha_{isi,0}\lambda_{isi,0}(1)+n_{g_{1},g_{2},g_{3}}(1),\dots,\alpha_{isi,0}\lambda_{isi,0}(K)+n_{g_{1},g_{2},g_{3}}(K)\}.
\ese
where {$n_{g_{1},g_{2},g_{3}}(k) = \sum_{s,t}1\{z_{isi,s,t}=k, v_{isi,s,t}=0, z_{isi,1,x_{1}}=g_{1},z_{isi,2,x_{2}}=g_{2},z_{isi,3,{(y_{t-1},y_t)}}=g_{3}\}$}.

\item Sample hyper-parameters $\alpha_{isi,0}$ and $\alpha_{isi}^{(0)}$ using the approximation method introduced in West (1992) for large sample size
\bse
&& \alpha_{isi,0} \mid \bzeta \sim \Ga\left\{a_{isi,0}+K-1, b_{isi,0}+\gamma+\log(n)\right\}, \\
&& \alpha_{isi}^{(0)} \mid \bzeta \sim \Ga\left\{a_{isi}^{(0)}+K-1, b_{isi}^{(0)}+\gamma+\log(n)\right\},
\ese 

where $n$ is the sample size 
and $\gamma$ is the Euler's constant.

\item Sample the mean probability vector $\blambda_{isi,0}$ as follows. 
First, for $\ell=1,\dots,n_{g_{1},g_{2},g_{3}}(k)$, we sample the auxiliary variable $\omega_{\ell}$ as
\bse
&& \omega_{\ell} \mid \bzeta \sim \Bern\left\{\frac{\alpha_{isi,0}\lambda_{isi,0}(k)}{\ell-1+\alpha_{isi,0}\lambda_{isi,0}(k)}\right\}.  
\ese
We set $m_{g_{1},g_{2},g_{3}}(k)=\sum_{\ell}\omega_{\ell}$, $m_{0}(k)=\sum_{g_{1},g_{2}.g_{3}}m_{g_{1},g_{2},g_{3}}(k)$. We then sample $\blambda_{isi,0}$ as
\bse
&& \blambda_{isi,0} \mid \bzeta \sim \Dir\{\alpha_{isi,00}/K+m_0(1),\dots,\alpha_{isi,00}/K+m_0(K)\}.
\ese

\item Sample the mixture gamma shape and rate parameters $\alpha_{k}$'s and $\beta_{k}$'s 

As discussed in Section 4 in the main paper, we use an approximation algorithm introduced in \cite{miller2019fast} to sample the shape parameters $\alpha_{k}$'s. 
Let $\alpha_{k}^{(i-1)}$ and $\beta_{k}^{(i-1)}$ be the shape and rate parameter from MCMC iteration $i-1$.
Initialize the parameters of the gamma conditional, $A_{k}$ and $B_{k}$, for $\alpha_{k}^{(i)}$ as follows:
\bse
&&\mu_{k} = \alpha_{k}^{(i-1)}/\beta_{k}^{(i-1)}, \nonumber \\
&&T_{k} = S_{k}/\mu_k-R_k+n_{k}\log(\mu_k)-n_{k}, \nonumber \\
&&A_{k} = a_{isi,0}+n_{k}/2, \nonumber \\
&&B_{k} = b_{isi,0}+T_{k}, \nonumber 
\ese
where $R_k=\sum_{z_{isi,s,t}=k} \log(\tau_{s,t})$, $S_{k}= \sum_{z_{isi,s,t}=k} \tau_{s,t}$ and $n_{k} = 1\{z_{isi,s,t}=k\}$.

Let $M$ be the maximum allowed number of iterations. 
We iteratively compute $A_{k}$ and $B_{k}$ until convergence.
For iteration $1,\dots,M$, compute the following: 
\bse
&& a_{k} = A_{k}/B_{k},\\
&& A_{k} = a_{isi,0}-na_{k}+na_{k}^2\psi'(a_{k}),\\
&& B_{k} = b_{isi,0}+(A_{k}-a_{isi,0})/a_{k}-n_{k}\log(a_{k})+n_{k}\psi(a_{k})+T_{k},
\ese
and return $A_{k}$ and $B_{k}$ if $|a_{k}/(A_{k}/B_{k})-1|<\epsilon$ for some predetermined tolerance $\epsilon$.
Note that $\psi(\cdot)$ and $\psi'(\cdot)$ represent the digamma and trigamma function, respectively.
Given $A_{k}$ and $B_{}$, the shape and rate parameter for the current iteration $i$, $\alpha_{k}^{(i)}$ and $\beta_{k}^{(i)}$, can be obtained via their gamma full conditionals
\bse
&& \alpha_{k}^{(i)} \sim \Ga(A_{k},B_{k}), \\
&& \beta_{k}^{(i)} \sim \Ga\left(1+\alpha_{k}^{(i)} n_{k}, 1+S_{k}\right).
\ese

\item Update the covariate cluster parameters $\bk_{isi}$ and $\bz_{isi,r}$'s

Recall $\bk_{isi}=(k_{isi,1},k_{isi,2},k_{isi,3})$ represents the number of clusters for each covariate $r$ and $\bz_{isi,r}=(z_{isi,r,1},\dots,z_{isi,r,d_{r}})$ indicate the cluster allocation for each level of covariate $r$.   
For each covariate $r$, values of $\bz_{isi,r}$ partition $d_{r}$ levels into $k_{isi,r}$ clusters.
The constructed clusters are denoted by $C_{isi}^{(r)}=\{\C_{isi,g_r}^{(r)}\}_{g_r=1}^{k_{isi,r}}$ for covariate $r$.
If $z_{isi,r,w}=g_r$, the $w\th$ level of covariate $r$ belongs to the $g_r\th$ cluster in $\C_{isi}^{(r)}$.
After integrating out $\blambda_{isi,g_{1},g_{2},g_{3}}$ and conditioning on the cluster configurations $\{\C_{isi}^{(r)}\}_{r=1}^{3}$, we have

\bse
&& p(\bz_{isi,r}\mid \{\C_{isi}^{(r)}\}, \bzeta) = \prod_{g_{1},g_{2},g_{3}}\frac{\beta\{\alpha_{isi,0}\lambda_{isi,0}(1)+n_{g_{1},g_{2},g_{3}}(1),\dots,\alpha_{isi,0}\lambda_{isi,0}(K)+n_{g_{1},g_{2},g_{3}}(K)\}}{\beta\{\alpha_{isi,0}\lambda_{isi,0}(1),\dots,\alpha_{isi,0}\lambda_{isi,0}(K)\}},
\ese

where $n_{g_{1},g_{2},g_{3}}(k) = \sum_{s,t} 1(z_{isi,s,t}=k,z_{isi,1,x_{1}}=g_{1},z_{isi,2,x_{2}}=g_{2},z_{isi,3,y_{t-1}}=g_{3})$.

To update the values of $\bk_{isi}$'s, we do the following.
If $k_{isi,r}<d_{r}$, we propose to increase $k_{isi,r}$ by 1 by randomly splitting a cluster into two. 
If $k_{isi,r}>1$, we propose to decrease $k_{isi,r}$ by 1 by randomly merging two clusters. 
We accept the new clustering if the log likelihood of the proposed move is greater than a sampled value from the log-uniform distribution. 
\end{enumerate}

In all our examples, we ran $10,000$ MCMC iterations with the first $2,000$ as burn-ins. 
Additionally, the rest of the samples are thinned by an interval of $5$.
MCMC diagnostic plots show no issues in mixing or convergence. 
The input of our program is a cleaned data set consisting of $70818$ rows and $6$ columns: mouse ID, genotype, social context, preceding syllable, current syllable, and ISI length.
We refer interested readers to \cite{sarkar2018bayesian} for details of the implementation of the model for the syllable transitions.
We provide \texttt{R} scripts for inference of the ISIs as part of the Supplemental Materials.

\section{Individual-Specific Mixture Coefficients}\label{sec:coeff}


{In our models for the transition probabilities and the ISIs, the coefficients $\bpi_{trans,0}^{(i)}$ for the population and individual-level effects are designed to be mouse-specific. 
This is in contrast to \cite{sarkar2018bayesian}, where they used a similar model but the coefficients $\bpi_{trans,0}$ did not vary among different individuals. 
Here, we display in Table~\ref{tab:coeff} the statistics of the coefficient $\bpi_{trans,0}^{(i)}$'s for transition probabilities for all 18 mice taken from the last MCMC iteration.}

\begin{table}[ht]
    \centering
    \begin{tabular}{ccccc}
         & min & max & mean & standard deviation  \\\hline
        \texttt{d} & $0.027$ & $0.693$ & $0.301$ & $0.198$ \\
        \texttt{m} & $0.028$ & $0.966$ & $0.518$ & $0.292$  \\
        \texttt{s} &$0.010$ & $0.488$ & $0.193$ & $0.152$  \\
        \texttt{u} & $0.009$ & $0.956$ & $0.383$ & $0.326$ \\\hline
    \end{tabular}
    \caption{Results for the Foxp2 data set showing the summary of coefficients $\bpi_{trans,0}^{(i)}$'s for transition probabilities given the preceding syllable. 
    The statistics are taken over the 18 mice from the last MCMC iteration. }
    \label{tab:coeff}
\end{table}

{We see that the coefficient $\pi_{trans,0}^{(i)}(y_{t-1})$'s differ substantially between different mice. 
This happens regardless of the preceding syllable $y_{t-1}$ but is especially prominent for the preceding syllable \texttt{u} 
for which the difference between the minimum and maximum coefficient is as high as $0.947$ and the standard deviation among all $18$ mice is $0.326$.
This discovery illustrates the utility of allowing the coefficients to be mouse-specific to allow better characterization of animal heterogeneity.}
The coefficients $\bpi_{isi,0}^{(i)}$ for ISIs are presented in Table~\ref{tab:coeff_isi}. 
{Unlike $\bpi_{trans,0}^{(i)}$'s,  the coefficients for ISIs, $\pi_{isi,0}^{(i)}(k)$'s,  come from a narrower range that is center around $0.5$.
This suggests that the individual- and the population-level effect that govern the transition dynamics appear to have a similar weight, regardless of the individual or the mixture component. }

\begin{table}[ht]
    \centering
    \begin{tabular}{ccccc}

         & min & max & mean & standard deviation  \\\hline
        Component 1 & $0.347$ & $0.629$ & $0.491$ & $0.069$ \\
        Component 2 & $0.277$ & $0.692$ & $0.486$ & $0.099$  \\
        Component 3 &$0.364$ & $0.599$ & $0.467$ & $0.055$  \\
        Component 4 & $0.387$ & $0.659$ & $0.509$ & $0.073$ \\\hline
    \end{tabular}
    \caption{Results for the Foxp2 data set showing the summary of coefficients $\bpi_{isi,0}^{(i)}$'s for ISIs given the mixture component. 
    The statistics are taken over the 18 mice from the last MCMC iteration. }
    \label{tab:coeff_isi}
\end{table}

\section{Simulation Studies} \label{sec: sim study}

We designed simulation experiments to evaluate the performance of the proposed MRP in assessing mouse vocalization dynamics.
We are not aware of any other model for vocalization syntax that can be directly compared with ours. 
We thus focus here on evaluating the performance of our proposed method in recovering an underlying `truth'.
Also, since the mixed Markov model was evaluated in detail in \cite{sarkar2018bayesian}, 
we focus mainly on evaluating the model for the ISIs here. 
In designing our simulation truths, we closely mimic the Foxp2 data set, 
including sampling the syllables $\{\texttt{d},\texttt{m},\texttt{s},\texttt{u}\}$ and the ISIs 
using the transition probabilities and the gamma mixtures estimated in Section~5 as the corresponding truths. 
We consider $18$ mice, $10$ with the Foxp2 mutation and $8$ wild-types, that sing under three social contexts, $\{U,L,A\}$.
The first syllable for each sequence is copied from the Foxp2 data set and the rest are sampled using the transition probabilities presented in Figure~4 in Section~5.
Given genotype, context and the preceding syllable, ISI is then sampled using the gamma mixture 
shown in Table~5 in Section~5.
We set the number of mixture components to be $K=4$ and analyze the following scenarios.


\begin{figure}[ht]
\centering
\includegraphics[width=.9\textwidth]{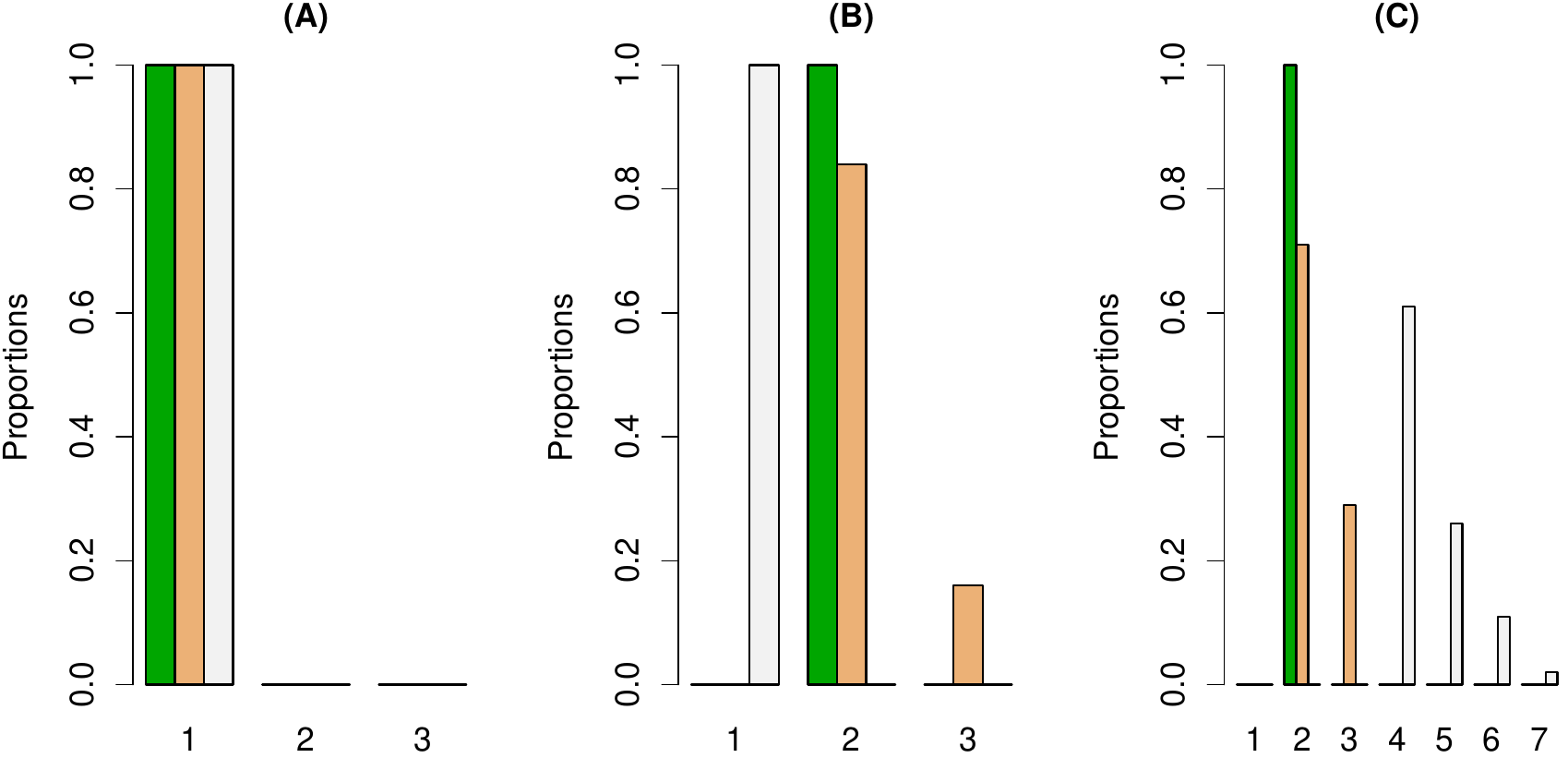}
\caption{Results of the simulation studies showing the percentages of the number of clusters $k_{isi,r}$ for each covariate $r$ among thinned samples after burn-ins for every study. 
The x-axis represents the number of clusters.
The green/orange/white bar represents the estimated posterior distribution of $P(k_{isi,r})$ for genotype $x_1 \in \{F,W\}$, social context $x_2 \in \{U,L,A\}$ and preceding-current syllable pair $(y_{t-1},y_t) \in \Y \times \Y = \{\texttt{d},\texttt{m},\texttt{s},\texttt{u}\} \times \{\texttt{d},\texttt{m},\texttt{s},\texttt{u}\}$, respectively. 
The results agree with the scenario settings for each simulation study.}
\label{fig:isi_sim_clusts}
\end{figure}

{
\begin{itemize}
\item [(A)] $\blambda_{isi, x_1,x_2,{(y_{t-1},y_t)}} =\hat{\blambda}_{isi,1,1,1}$ for $x_1=1,2$, $x_2=1,2,3$, $y_{t-1}=1,2,3,4$,  {and $y_t = 1,2,3,4$}.
ISI does not vary with genotype, context or the {preceding-current syllable pair} ($k_{isi,1,0}=1,k_{isi,2,0}=1,k_{isi,3,0}=1$).
\item [(B)]  $\blambda_{isi, x_1,x_2,{(y_{t-1},y_t)}} =\hat{\blambda}_{isi,x_1,x_2,1}$ for $x_1=1,2$, $x_2=1,2,3$, $y_{t-1}=1,2,3,4$,  {and $y_t = 1,2,3,4$}.
ISI varies with genotype and context but does not vary with the {preceding-current syllable pair} ($k_{isi,1,0}=2,k_{isi,2,0}=3,k_{isi,3,0}=1$).
\item [(C)] $\blambda_{isi, x_1,x_2,{(y_{t-1},y_t)}} =\hat{\blambda}_{isi,x_1,x_2,{(y_{t-1},y_t)}}$ for $x_1=1,2$, $x_2=1,2,3$, $y_{t-1}=1,2,3,4$,  {and $y_t = 1,2,3,4$}.
ISI varies with genotype, context and the {preceding-current syllable pair} ($k_{isi,1,0}=2,k_{isi,2,0}=3,k_{isi,3,0}={16}$).
\end{itemize}
}

Figure~\ref{fig:isi_sim_clusts} displays the estimated posterior probability $\wh{P}(k_{isi,r}  \mid \data)$ for each covariate $r$.  
{We see that $\wh{P}(k_{isi,r}=k_{isi,r,0} \mid \data)\approx 1$ for genotype across the three studies, 
showing great performance of our method in testing the relevant hypotheses. 
When we let ISI vary with the context, i.e.,  in studies (B) and (C),  the number of clusters for  the context splits between two and three, matching the result we have seen for the Foxp2 data set in Figure~7. } 

\begin{figure}[!ht]
\centering
\includegraphics[width=\textwidth]{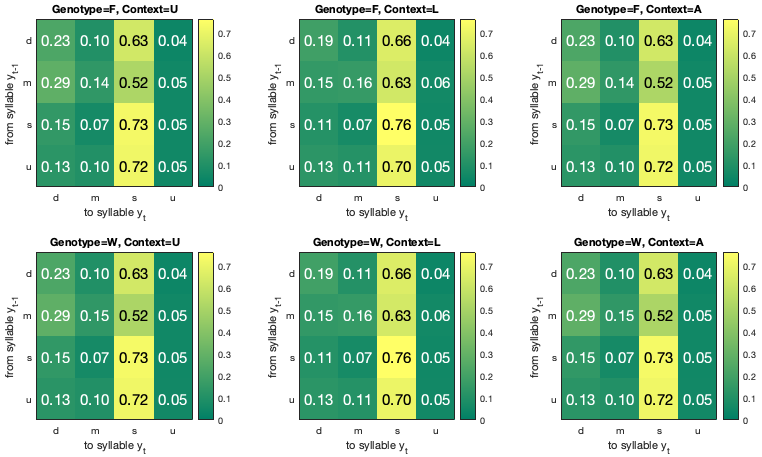}
\caption{Results for transitions for simulation design (C) showing the estimated posterior mean of the transition probabilities $P_{trans,x_1,x_2}(y_t \mid y_{t-1})$ of syllables $y_t, y_{t-1} \in \Y = \{\texttt{d},\texttt{m},\texttt{s},\texttt{u}\}$ under different combinations of genotype $x_1\in \{F,W\}$ and social context $x_2\in\{U,L,A\}$.}
\label{fig: trans_mean_sim}
\end{figure}

\begin{figure}[!ht]
\centering
\subfloat[Histogram of the transformed ISIs with the estimated posterior mean (red line) of their  marginal gamma mixture density based on MCMC samples after burn-in and thinning.]{
\includegraphics[width=0.43\textwidth]{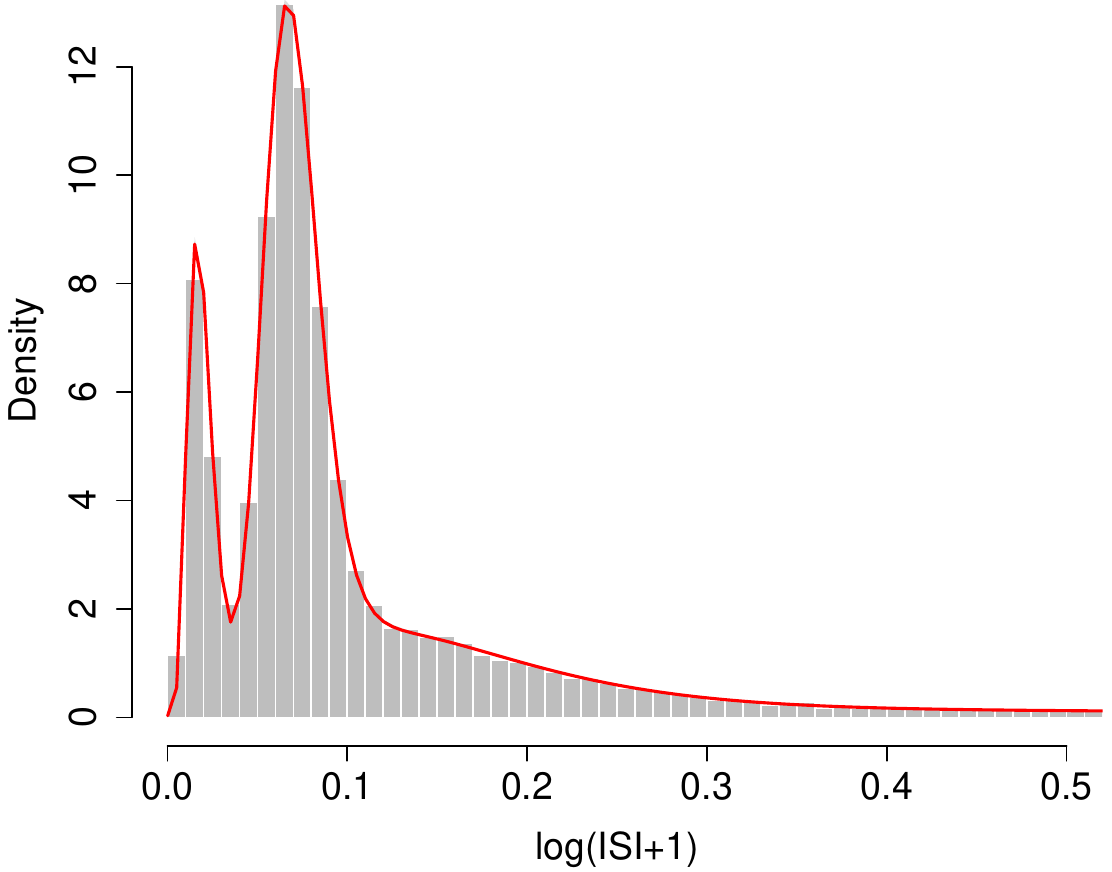}
\label{fig:isi_hist_total_sim}}
\qquad
\subfloat[Histograms of the transformed ISIs for each component of the gamma mixture model along with the component density (red lines) from the last MCMC iteration. The x-axes are adjusted for better visualization.]{
\includegraphics[width=0.47\textwidth]{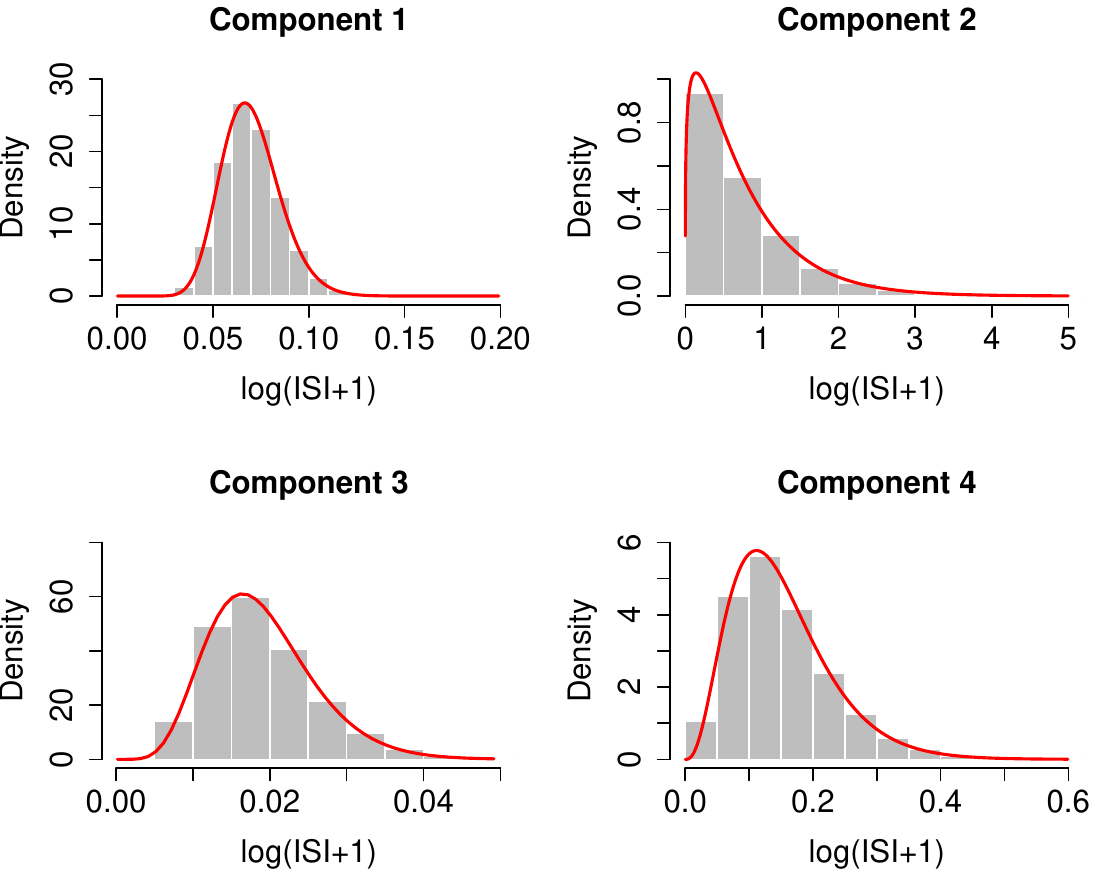}
\label{fig:isi_hist_comp_sim}}
\caption{Results for ISIs for simulation design (C).
 }
\label{fig:isi_hist_post_sim}
\end{figure}

Study (C) is the closest to the Foxp2 data set where all of the covariates are significant for the ISIs, and we now compare the results from study (C) to those of the Foxp2 data set. 
We first look at the estimated posterior mean for transition probabilities under every combination of genotype and social context, presented in Figure~\ref{fig: trans_mean_sim}.
For every transition type, given genotype and social context, the estimated transition probability is very close to that of the Foxp2 data set (Figure~4). 
%
Next, we turn to the results for ISIs. 
Figure~\ref{fig:isi_hist_total_sim} displays the histogram of the simulated data superimposed over the estimated posterior mean ISI distribution. 
The shape of the synthesized data looks close to the real data set (Figure~6) and the estimated posterior density fits the data well.
The histogram of ISIs for each component of the mixture gamma is presented in Figure~\ref{fig:isi_hist_comp_sim}
along with their estimates from the last MCMC iteration. 
We observe that there is no label switching.  
%
The estimated values of gamma mixture parameters are presented in Table~\ref{tab:isi_heatmap_sim}. 
The shape and the rate parameters are both similar to the corresponding estimated values for the Foxp2 data set (Table~5 from the main article). 
Similarly, the mixture probabilities for each covariate also agree with the corresponding simulation truths. 
The performance of our method in all of the different simulation scenarios, 
including scenario (C) where the simulation design closely matched the real data results in Section~5,  
shows that our model is capable of assessing the mouse vocalization dynamics under a wide range of realistic scenarios.

\begin{table}[ht]
\parbox{.29\linewidth}{
\centering
\begin{tabular}{ccc}
&shape.k & rate.k  \\ \cline{2-3}
Comp 1 & $21.10$                   & $301.59$         	 \\
Comp 2 & $1.23$                     & $1.65$          	\\
Comp 3 & $7.50$                 & $395.39$           \\
Comp 4 &$3.77$                   & $24.85$          \\\cline{2-3}
\end{tabular}
\caption*{{Estimated gamma shape and rate parameters}}
}
~~~
\parbox{.29\linewidth}{
\centering
\begin{tabular}{ccc}
&$F$ & $W$  \\ \cline{2-3}
Comp 1 & $0.53$                   & $0.57$         	 \\
Comp 2 &  $0.16$                     & $0.08$          	\\
Comp 3 &  $0.09$                 & $0.11$           \\
Comp 4 & $0.21$                   & $0.24$          \\\cline{2-3}
\end{tabular}
\caption*{{Estimated mixture probabilities for genotypes}}
}
~~
\parbox{.29\linewidth}{
\centering
\begin{tabular}{cccc}
&$U$ & $L$ & $A$  \\ \cline{2-4}
Comp 1 & $0.60$                   & $0.46$       & $0.60$   	 \\
Comp 2&  $0.12$                     & $0.12$       & $0.12$      	\\
Comp 3&  $0.06$                 & $0.18$         & $0.06$     \\
Comp 4 & $0.22$                   & $0.24$        & $0.22$     \\\cline{2-4}
\end{tabular}
\caption*{{Estimated mixture probabilities for contexts}}
} 
\newline
\vspace*{5pt}
\newline
\parbox{\linewidth}{
\centering
\begin{tabular}{ccccccccc}
&$(\ttd,\ttd)$ & $(\ttd,\ttm)$ & $(\ttd,\tts)$ & $(\ttd,\ttu)$ &$(\ttm,\ttd)$ & $(\ttm,\ttm)$ & $(\ttm,\tts)$ & $(\ttm,\ttu)$ \\ \cline{2-9}
Comp 1 & 0.62 &0.62& 0.53& 0.50& 0.62 &0.62& 0.50& 0.62   	 \\
Comp 2&  0.11 &0.11 &0.13& 0.12 &0.11& 0.11& 0.12 &0.11      	\\
Comp 3&  0.06 &0.06 &0.09& 0.16& 0.06& 0.06 &0.16 &0.06     \\
Comp 4 & 0.21 &0.21& 0.25 &0.22& 0.21 &0.21 &0.22 &0.21 \\\cline{2-9}
\end{tabular}
\caption*{{Estimated mixture probabilities for each preceding-current syllable pair}}
}
\newline
\vspace*{5pt}
\newline
\parbox{\linewidth}{
\centering
\begin{tabular}{ccccccccc}
&$(\tts,\ttd)$ & $(\tts,\ttm)$ & $(\tts,\tts)$ & $(\tts,\ttu)$ &$(\ttu,\ttd)$ & $(\ttu,\ttm)$ & $(\ttu,\tts)$ & $(\ttu,\ttu)$\\ \cline{2-9}
Comp 1 & 0.53 &  0.62  & 0.41  & 0.62  & 0.50 & 0.53  & 0.50 &  0.53	 \\
Comp 2&  0.13  & 0.11  & 0.21  & 0.11  & 0.12&  0.13&   0.12 &  0.13  	\\
Comp 3& 0.09  & 0.06  & 0.11  & 0.06  & 0.16&  0.09&   0.16 &  0.09 \\
Comp 4 &  0.25  & 0.21  & 0.27  & 0.21  & 0.22&  0.25 &  0.22 &  0.25  \\\cline{2-9}
\end{tabular}
\caption*{{Estimated mixture probabilities for each preceding-current syllable pair (cont'd)}}
}
\caption{{Results for ISIs for simulation design (C) taken from the last MCMC iteration.}}
\label{tab:isi_heatmap_sim}
\end{table}

\clearpage
\section*{References}






\refmark
Geisser, S. and Eddy, W. F. (1979). A predictive approach to model selection. \emph{Journal of the American Statistical Association}, {\bf 74}, 153–160.

\refmark
Miller, J. W. (2019). Fast and accurate approximation of the full conditional for gamma shape parameters. 
\emph{Journal of Computational and Graphical Statistics}, {\bf 28}, 476-480.


\refmark
Sarkar, A., Chabout, J., Macopson, J. J., Jarvis, E. D., and Dunson, D. B. (2018). Bayesian semiparametric mixed effects Markov models with application to vocalization syntax.
\emph{Journal of the American Statistical Association}, {\bf 113}, 1515-1527.

\refmark
Teh, Y. W., Jordan, M. I., Beal, M. J., and Blei, D. M. (2006). Hierarchical Dirichlet processes. 
\emph{Journal of the American Statistical Association}, {\bf 101}, 1566-1581.


\refmark
Watanabe, S. (2010). Asymptotic equivalence of Bayes cross validation and widely applicable information criterion in singular learning theory. \emph{Journal of Machine Learning Research}, {\bf 11}, 3571–3594.

\refmark 
West, M. (1992). Hyperparameter estimation in Dirichlet process mixture models. Institute of Statistics and Decision Sciences, Duke University, Durham, USA, Technical report.


\end{document}